\documentclass[aps,prd,twocolumn,superscriptaddress,nofootinbib,showpacs,letterpaper,eqsecnum]{revtex4-1}

\usepackage{mathptmx}
\usepackage{showlabels}
\usepackage{verbatim}
\usepackage[usenames,dvipsnames]{color}
\usepackage[pdftex]{graphicx}
\usepackage[english]{babel}
\usepackage{amsmath,amssymb}
\usepackage[colorlinks=true,citecolor=blue]{hyperref}

\begin{document}

\title{Coupled Oscillator Model for Nonlinear Gravitational Perturbations}

\author{Huan Yang}
\affiliation{Perimeter Institute for Theoretical Physics, Waterloo, Ontario N2L 2Y5, Canada}
\affiliation{Institute for Quantum Computing, University of Waterloo, Waterloo, Ontario N2L3G1, Canada}
\author{Fan Zhang}
\affiliation{Center for Cosmology and Gravitational Wave, Department of Astronomy, Beijing Normal University, Beijing 100875, China}
\affiliation{\mbox{Department of Physics, West Virginia University, PO Box 6315, Morgantown, WV 26506, USA}}
\author{Stephen R. Green}
\affiliation{Perimeter Institute for Theoretical Physics, Waterloo, Ontario N2L 2Y5, Canada}
\author{Luis Lehner}
\affiliation{Perimeter Institute for Theoretical Physics, Waterloo, Ontario N2L 2Y5, Canada}
\affiliation{CIFAR, Cosmology \& Gravity Program, Toronto, ON M5G 1Z8, Canada}

\begin{abstract}
  Motivated by the gravity/fluid correspondence, we introduce a new
  method for characterizing nonlinear gravitational
  interactions. Namely we map the nonlinear perturbative form of the
  Einstein equation to the equations of motion of a collection of
  nonlinearly-coupled harmonic oscillators. These oscillators
  correspond to the quasinormal or normal modes of the background
  spacetime.  We demonstrate the mechanics and the utility of this
  formalism within the context of perturbed asymptotically anti-de
  Sitter black brane spacetimes.  We confirm in this case that the
  boundary fluid dynamics are equivalent to those of the hydrodynamic
  quasinormal modes of the bulk spacetime. We expect this formalism to
  remain valid in more general spacetimes, including those without a
  fluid dual. In other words, although borne out of the gravity/fluid
  correspondence, the formalism is fully independent and it has a much
  wider range of applicability. In particular, as this formalism
  inspires an especially transparent physical intuition, we expect its
  introduction to simplify the often highly technical analytical
  exploration of nonlinear gravitational dynamics.
\end{abstract}


\maketitle

\section{Introduction}

Can spacetimes become turbulent?  Direct numerical simulations of
large asymptotically anti--de Sitter (AdS) black
holes~\cite{Adams:2013vsa} and their holographically dual
fluids~\cite{Carrasco:2012nf,Green:2013zba} have provided convincing
evidence that this is the case. This phenomenon, perhaps
counterintuitive at first 
glance,\footnote{Due to a crucial difference: the Einstein equation is
  linearly degenerate as opposed to truly nonlinear as is the case of
  e.g., the Navier-Stokes equations.} can be understood through the
gravity/fluid
correspondence~\cite{Baier:2007ix,Bhattacharyya:2008jc,VanRaamsdonk:2008fp}.
This correspondence links the behavior of long-wavelength
perturbations of black holes in AdS to viscous relativistic
hydrodynamics, and its regime of applicability can include cases of
high Reynolds number on the fluid side. Spacetime turbulence then
follows from turbulence in the dual
fluid~\cite{VanRaamsdonk:2008fp,Carrasco:2012nf}.  On the gravity
side, a high Reynolds number corresponds to dissipation of
gravitational perturbations that is weak when compared with nonlinear
interactions.  It is therefore not surprising that it arises in the
vicinity of asymptotically AdS black holes, which can have relatively
long lived quasinormal modes.

The observation of gravitational turbulence in AdS motivates a further
question: Can one analyze this striking nonlinear behavior directly in
general relativity {\em without relying on the existence of a
  holographic dual}? That is, rather than borrowing from the dual
hydrodynamic description---and any restricted regime of
applicability---can one establish a bona-fide description of
turbulence as a perturbative solution of the Einstein equation?
Recall that turbulence is a nonlinear phenomenon characterized, in
particular, by cascades of energy (and sometimes enstrophy) between
wave numbers.  It is therefore delicate to fully capture this behavior
within ordinary perturbation theory without carrying it out to
sufficiently high orders and performing a suitable
resummation~\cite{Green:2013zba}.  In order to take into account the
essential gravitational self-interactions of perturbations that are
present in the Einstein equation we will require a more general
perturbative framework.

In this work we introduce a nonlinear coupled-oscillator model to
describe the {\em interaction} of quasinormal or normal modes of a
background spacetime, in particular their mode-mode couplings.  This
proposal is a natural generalization of our earlier study of nonlinear
scalar wave generation around rapidly-spinning asymptotically flat
black holes \cite{Yang:2014tla}, where the back-reaction on the
driving mode was neglected (we account for it properly in this paper).
This previous model illustrated that the onset of turbulence in
gravity does not require the spacetime to be asymptotically anti--de
Sitter\footnote{In analogy to hydrodynamics, it is of course necessary
  to be in the regime of high {\em gravitational Reynolds
    number}.}. In the nonlinear oscillator model presented here, the
coupling between modes is accounted for explicitly and in real time as
opposed to implicitly through a recursive scheme. Therefore the
equations of motion provide solutions that are valid over longer time
scales.

Within this model, nonlinear gravitational perturbations are described
by excitations of modes (quasinormal or normal).  For a given
background spacetime, the collection of modes is parametrized by a
particular set of frequencies, damping rates, and, at the nonlinear
level, {\em mode coupling coefficients}.  Through these parameters, we
can quantitatively compare and contrast signatures of nonlinear
gravitational perturbations in different backgrounds, in the same way
that frequencies and damping rates alone characterize linear
perturbations.  In this way we can gain a better understanding of
nonlinear interactions and associated phenomena (such as turbulence)
in general relativity. The route taken when constructing this
formalism essentially offers a new perspective on how to deal with
nonlinear metric perturbations that is conducive to intuition
building. This compares favorably with more traditional methods, where
one has to contend with difficult technical details that often mask
the underlying physics.

To provide a concrete example, we will apply our methods to study
nonlinear perturbations of an asymptotically AdS black brane.  The
gravity/fluid correspondence applies in this case and the resulting
coupled-oscillator system may be compared against the dual fluid.  We
find that our equations are consistent with the relativistic
hydrodynamic equations provided by the duality. Although the agreement
is expected, our derivation provides an explicit demonstration and a
natural physical interpretation of the observed phenomena in terms of
quasinormal modes. We emphasize that the derivations in the gravity
and fluid sides are independent of each other, and so the treatment
for gravitational perturbations does not depend on the existence of a
dual fluid and can be applied to more general spacetimes.

In the interest of caution, we recall that quasinormal modes do not
form a basis for generic metric perturbations
(see~\cite{Warnick:2013hba} for a recent discussion).  For instance,
consider linear perturbations of the (asymptotically flat) Kerr
spacetime as an example (see also discussions in Sec. \ref{sec2}). The
signal sourced by some matter distribution comprises quasinormal
modes, the late-time ``tail'' term, as well as a prompt piece that
travels along the light cone. In this sense, our formalism is
approximate as we consider only the quasinormal mode
contributions. However, in many cases, such as the ringdown stage of
binary black hole mergers or when considering long wavelength
perturbations of an asymptotically AdS black brane, it is sufficient
to track only the quasinormal modes, as they are the dominant part of
the signal (see, e.g.,~\cite{Barranco:2013rua}, for a related
discussion). In more general scenarios, we can always check the
validity of our approximation by estimating the magnitudes of the
other contributions.

This paper is organized as follows. In Sec.~\ref{sec2}, we introduce
the general formalism of the nonlinear coupled-oscillator model, and
compare it with traditional methods for handling nonlinear
gravitational perturbations. In Sec.~\ref{sec3}, we briefly review the
asymptotically AdS black brane spacetime and the gravity/fluid
correspondence, and we analyze the boundary fluid in the
mode-expansion picture. In Sec.~\ref{sec4}, we apply the general
formalism to the specific case of the asymptotically AdS black brane.
We conclude in Sec.~\ref{conclusion}. The gravitational constant $G$
and the speed of light $c$ are both set to one, unless otherwise
specified.  Appendices are provided to elaborate on certain details.

\section{General Formalism}\label{sec2}

In this section, we begin by reviewing the traditional approach to solving
the Einstein equation using ordinary perturbation theory and assuming
a series expansion in the perturbation amplitude.  This method might
not lend itself to easily capturing relevant phenomena like
turbulence.  In the case where the linearized dynamics take the form of
independently evolving normal or quasinormal modes (in the absence or
presence of dissipation, respectively), we then show how the nonlinear
Einstein equation can be represented as a set of coupled oscillator
equations, which is analogous to treatments of the Navier-Stokes
equation in fluid dynamics, and {\em is} indeed capable of cleanly capturing
turbulence.  For simplicity, we restrict our discussion to vacuum
spacetimes, but it is straightforward to generalize the analysis to
spacetimes with a cosmological constant.

\subsection{Ordinary perturbation theory}

Given any metric $g_{\mu\nu}$, one can split it into the sum of a
``background'' metric and a ``perturbation'',
\begin{equation}\label{eqdefh}
g_{\mu\nu} = g^{\rm B}_{\mu\nu}+h_{\mu\nu}\,.
\end{equation}
Without invoking any approximation, the vacuum Einstein equation may then be written as
\begin{equation}\label{eqpertein}
R_{\mu\nu}(g^{\rm B}) + R^{(1)}_{\mu\nu}(g^{\rm B},h) + R^{(2)}_{\mu\nu}(g^{\rm B},h) +\sum_{n=3}^\infty R^{(n)}_{\mu\nu}(g^{\rm B},h)=0\,,
\end{equation}
where $R^{(n)}_{\mu\nu}(g^{\rm B},h)$ denotes the $n$th order Ricci
tensor expanded about $g^{\rm B}_{\mu\nu}$.  Explicitly, the
linearized and second order terms are
\begin{equation}
R^{(1)}_{\mu\nu} \equiv \frac{1}{2} (-h_{|\mu\nu} -{h_{\mu\nu|\alpha}}^\alpha+{h_{\alpha \mu|\nu}}^\alpha + {h_{\alpha \nu | \mu}}^\alpha)\,,
\end{equation}
and
\begin{align}\label{eqr2}
R^{(2)}_{\mu\nu} \equiv &
\frac{1}{4} \left[ 
h^{\alpha\beta}{}_{|\nu} h_{\alpha\beta|\mu} 
+2\left( h_{\nu\alpha|\beta} - h_{\nu \beta|\alpha} \right)h_{\mu}{}^{\alpha|\beta}  \right . \nonumber  \\
&\left . 
+ \left( h_{\alpha \mu |\nu} + h_{\alpha \nu|\mu} - h_{\mu \nu |\alpha}\right)\left(h_{\beta}{}^{\beta|\alpha} - 2 h^{\alpha \beta}{}_{|\beta} \right)
\notag \right. \\ & \left. 
+2h^{\alpha\beta}\left( h_{\alpha \beta|\mu|\nu} + h_{\mu \nu|\alpha |\beta} - h_{\alpha\mu | \nu|\beta} - h_{\alpha\nu | \mu|\beta} \right)
\right]\,.
\end{align}
In these expressions, covariant derivatives associated to the
background metric $g^{\rm B}_{\mu\nu}$ are denoted by vertical lines.
The background metric is also used to raise and lower indices.

As described in~\cite{Wald}, ordinary perturbation theory 
assumes the existence of a one-parameter family of solutions
$g_{\mu\nu}(\epsilon)$, where $g_{\mu\nu}(0)=g_{\mu\nu}^{\text B}$, and
$h_{\mu\nu}$ depends differentiably on $\epsilon$.  One can then
Taylor expand the perturbation,
\begin{equation}\label{eq:hexpansion}
  h_{\mu\nu} = \epsilon h^{(1)}_{\mu\nu}+\epsilon^2 h^{(2)}_{\mu\nu} + \cdots.
\end{equation}
Perturbative equations of motion of order $n$ follow by
differentiating the Einstein equation~\eqref{eqpertein} $n$ times with
respect to $\epsilon$, and then setting $\epsilon=0$.  At zeroth order
we have simply
\begin{equation}
  R_{\mu\nu}(g^{\text B}) = 0\,,
\end{equation}
so that $g^{\text B}_{\mu\nu}$ is a vacuum solution itself.

At first order in $\epsilon$ we have the linearized Einstein equation,
\begin{equation}
R^{(1)}_{\mu\nu}(g^{\text B},h^{(1)}) =0\,.
\end{equation}
It is generally much easier to solve this equation (after making
appropriate gauge choices and imposing boundary and initial
conditions) than it is to solve the full Einstein equation.  Then for
sufficiently small $\epsilon$, $g_{\mu\nu}^{\text B} + \epsilon
h_{\mu\nu}^{(1)}$ should be a good approximation to
$g_{\mu\nu}(\epsilon)$.

This procedure may be continued to higher orders.  For instance, at second order, we obtain
\begin{equation}
 R^{(1)}_{\mu\nu} (g^{\text B},h^{(2)}) =- R^{(2)}_{\mu\nu}(g^{\text B},h^{(1)})\,.
\end{equation}
The second order perturbation is seen to evolve in the background
spacetime $g^{\text{B}}_{\mu\nu}$, and it is sourced by the first
order solution $h^{(1)}_{\mu\nu}$.

Generically, this approach reduces the nonlinear problem to a series
of linear inhomogeneous problems of the form
\begin{equation}\label{eq:PTschem}
  R^{(1)}_{\mu\nu}(g^{\text B},h^{(n)}) = S_{\mu\nu}^{(n)}(g^{\text B};h^{(1)},\ldots,h^{(n-1)})\,.
\end{equation}
Thus, at each order, one solves a linear partial differential equation
with a source, subject to appropriate boundary conditions and gauge
choices.  The left hand side of the equation at order $n$ consists
always of the $n$th order perturbation $h^{(n)}_{\mu\nu}$ evolving
linearly in the background spacetime $g^{\text{B}}_{\mu\nu}$.  The
source term $S_{\mu\nu}^{(n)}$ involves only already-solved lower
order pieces $h_{\mu\nu}^{(m)}$ for $m<n$, so a higher order
perturbation does not backreact on one of lower order.  Moreover,
since the $n$th order perturbation evolves in the zeroth order
background metric---not the $(n-1)$th order metric---the efficient
capture of parametric resonance type effects is
precluded~\cite{Green:2013zba,Yang:2014tla}.  (Of course, with enough
intuition, it may be possible to identify this behavior through a
suitable resummation of perturbations of sufficiently high order.)  In
following this program, the calculations are quite involved and the
gauge choices at different orders are often subtle (see
e.g.,~\cite{Bruni:1996im,Gleiser:1998rw,Ioka:2007ak,Brizuela:2009qd}). In
the specific context of extreme mass ratio binaries, recent examples
of this program are given in~\cite{Pound:2012nt,Gralla:2012db}. 

\subsection{Larger perturbations}\label{sec:larger}

After iterating the above procedure to any given order, the resulting
perturbative metric should be a good approximation to
$g_{ab}(\epsilon)$ {\em for sufficiently small} $\epsilon$.  However,
in certain situations one may be interested in studying systems with
{\em larger} (but still small) values of $\epsilon$, where the Taylor
expansion~\eqref{eq:hexpansion} either fails to converge or would
require a large number of terms to obtain a good solution.
Typically the perturbative solution would be valid for a short time,
but for long times secular terms might dominate. Therefore, a more
suitable scheme would be required.  In, for example, the
context of the Navier-Stokes equation, ordinary perturbation theory
might be capable of capturing the initial onset of turbulence, but it
would be ineffective in capturing fully developed
turbulence (and likewise for gravitational turbulence~\cite{Green:2013zba,Yang:2014tla}).

In order to characterize the nonlinear dynamics in general relativity
in a more efficient and transparent manner, we present here an
alternative way of obtaining approximate solutions that is better
suited for exploring certain nonlinear phenomena such as wave
interactions and turbulence.  We assume as before that
$g_{\mu\nu}^{\text B}$ satisfies the vacuum Einstein equation.  But
then, rather than Taylor expanding $h_{\mu\nu}$ as
in~\eqref{eq:hexpansion}, we consider the full metric perturbation
$h_{\mu\nu}$, and we attempt to solve directly a truncated version
of~\eqref{eqpertein}.  In fact truncation at second order,
\begin{equation}\label{eq:htruncated}
  R^{(1)}_{\mu\nu}(g^{\text B},h) + R^{(2)}_{\mu\nu}(g^{\text B},h) = 0\,,
\end{equation}
captures the essential nonlinearities of interest to us here.  We note
that our formalism could straightforwardly be extended to higher
orders, but for simplicity we restrict to second order nonlinearities
here.

To summarize, instead of solving a tower of inhomogeneous linear
equations \eqref{eq:PTschem} we solve a {\em nonlinear equation}, but we
neglect the higher order nonlinearities.  Instead of dealing with gauge issues at each order, we have only to impose the gauge condition
once on $h_{\mu\nu}$.  Of course, the truncation of the Ricci tensor
is not a tensor itself so the equation~(\ref{eq:htruncated}) is not gauge
invariant.  But it should be sufficient to the order we are working ($O(h^2)$).
As we shall see, this approach readily captures the nonlinear
mode coupling effects of interest to us.

In general it will be very difficult to solve~\eqref{eq:htruncated},
even neglecting the higher order nonlinearities as we have done.
However, as we describe in the following subsection, in cases where
the linear dynamics is dominated by the evolution of normal or
quasinormal modes,~\eqref{eq:htruncated} reduces to a system of
nonlinearly coupled (and possibly damped) oscillators.

\subsection{Expansion into modes}\label{sec:ModeExp}

We now restrict consideration to background spacetimes whose {\em linear}
perturbations are characterized (for some region of spacetime) by a
set of modes (normal or quasinormal).  In this case the first order
metric perturbation may be written
\begin{equation}\label{eqhdec-lin}
  h_{\mu\nu}^{(1)} (t, {\bf x}) \sim \sum_j [ q^-_j(t) \mathcal{Z}^{(j-)}_{\mu\nu}({\bf x})  +q^+_j(t) \mathcal{Z}^{(j+)}_{\mu\nu}({\bf x})]\,,
\end{equation}
with
\begin{equation}\label{eqqab-const}
q^-_j(t) = A_j e^{-i\omega_j t},\quad q^+_j(t)= B_j e^{i \omega^*_j t}\,.
\end{equation}
The background spacetime is assumed to be stationary and the $t$
coordinate is the associated Killing parameter.  Modes always occur in
pairs with frequencies $\omega_j$ and $-\omega_j^\ast$, so we have
organized the summation above along these lines, labeling each pair
with a multi-index $j$ (denoting both the transverse harmonic and
radial overtone).  The associated spatial wave functions are denoted
$\mathcal{Z}^{(j\pm)}_{\mu\nu}({\bf x})$.  Finally, $q^\pm_j$ and
$\{A_j,B_j\}$ are the displacements and the amplitudes for modes
$j\pm$, respectively.  As $h_{\mu\nu}$ must be real at all time, we expect that
$\{A_j,B_j\}$ (as well as
$ \{\mathcal{Z}^{(j-)}, \mathcal{Z}^{(j+)}\}$) are conjugate to each
other. 

The reason we organize our modes into pairs in~\eqref{eqhdec-lin} is
to emphasize that {\em all} modes must be included in the nonlinear
analysis; many linear analyses use symmetry arguments to only treat
modes with $\Re(\omega)>0$ \cite{Berti2009}.  In the case of
{\em normal} modes, the mode functions $\mathcal{Z}^{(j\pm)}_{\mu\nu}$
are degenerate and $\omega_j\in\mathbb{R}$, so we take
$q_j = q^-_j+q^+_j = A_je^{-i\omega_jt}+B_je^{i\omega_jt}$.  For {\em
  quasinormal} modes, the radial dependence of $\mathcal{Z}^{(j\pm)}_{\mu\nu}$, along
with the dissipative boundary conditions at the horizon and/or infinity, fixes the
time dependence of the mode uniquely.  Any ``degenerate'' mode in this
case must therefore have $\omega_j=-\omega_j^\ast$, so the frequency
is purely imaginary, and the multi-index $j$ describes just a single
mode.  We analyze these cases separately from the non-degenerate case
in the following sections.

Frequencies of quasinormal modes have nonzero positive imaginary part,
which implies an exponential time decay as a result of energy
dissipation.  In addition, this complex frequency means that the mode
functions generally blow up at spatial infinity and the horizon
bifurcation surface. However, as physical observers effectively lie
near null infinity, the quasinormal-mode signals they observe are
finite and the modes are indeed physical perturbations of the
spacetime.  For such observers, the sum in~\eqref{eqhdec-lin} can
become a good approximation over finite time intervals, although we
remind the reader that quasinormal modes do not form a complete basis
for generic metric perturbations\footnote{This qualification is
  represented by the use of the ``$\sim$'' notation
  in~\eqref{eqhdec-lin} (see, e.g.,~\cite{Kokkotas1999}).}.
Additional contributions to the metric can arise at late times from
waves being scattered by the background potential at large distances
(the ``tail'' term), or at early times from a prompt signal (on the
light cone) from the source (see,
e.g.,~\cite{Leaver1986,Kokkotas1999,Berti2009,Casals:2013mpa}); we
collect these into the ``residual part''. 

In this paper our focus is on mode-mode interactions and the
associated coupling coefficients.  We will therefore not consider the
nonlinear interactions between the modes and the tail and prompt
components of the metric perturbation. We caution, however, that such
couplings need not always be small.  While they are small for
perturbations of AdS black branes in the hydrodynamic limit (which we
analyze below), readers should keep in mind that they will lead to
additional contributions to, e.g., Eq.~\eqref{eq:pluggedin} below.
Furthermore, questions as to how quasinormal modes are excited by
moving matter, or how to compute the excitation factors for these
modes based on some arbitrary initial data are also beyond the scope
of this work (see \cite{Leaver1986, Hadar09, ZhangZhongYang2013} and Appendix
\ref{sec:schwarz}).

With these observations in mind, following the discussion in
Sec.~\ref{sec:larger} we write the {\em full} metric perturbation as
\begin{eqnarray}\label{eqhdec}
 h_{\mu\nu} (t, {\bf x}) &=&  \sum_j [ q^-_j(t) \mathcal{Z}^{(j-)}_{\mu\nu}({\bf x}) + q^+_j(t) \mathcal{Z}^{(j+)}_{\mu\nu}({\bf x}) ]\nonumber \\
 &&\quad+\text{ residual part,}
\end{eqnarray}
but now generalizing the coefficients $A_j$ and $B_j$ to be functions of time,
\begin{equation}\label{eqqab}
 q^-_j(t) = A_j(t) e^{-i\omega_j t},\quad q^+_j(t)= B_j(t) e^{i \omega^*_j t}\,.
\end{equation}
Our task is to determine the nonlinear evolution of quasinormal modes;
in other words, to evaluate the time dependence of
$q^{\pm}_j$. Addressing this task is generally nontrivial as it
requires the proper separation of the quasinormal modes from the
residual part of the full metric perturbation. For Schwarzschild and
Kerr spacetimes this is achievable by invoking the Green's function
technique (Appendix \ref{sec:schwarz}), whereas the generalization of
this approach to generic spacetimes remains an open problem. To present the coupled-oscillator model, we apply an alternative strategy of plugging~\eqref{eqhdec}
into the truncated Einstein equation~\eqref{eq:htruncated} and
projecting our the spatial dependencies, thereby obtaining mode
evolution equations. This method is most accurate for dealing with normal-mode evolutions and cases where the residual parts are negligible (for example, see Sec.~\ref{sec4}). In more general scenarios, we shall make several additional
approximations (such as neglecting certain time derivatives,
neglecting the residual part) to single out the ordinary differential
equations for $q^{\pm}_j$. We also caution that since the set of modes
generally does not form a complete basis, the resulting $h_{\mu\nu}$
is still only an approximate solution to the truncated Einstein
equation. For simplicity, hereafter we shall not explicitly write down
the residual part in the equations.

Upon substitution, the truncated Einstein
equation~\eqref{eq:htruncated} takes the form 
\begin{eqnarray}\label{eq:pluggedin}
  &&\sum_j\sum_{s=\pm}\left[ \rho_j^s({\bf x}) \ddot{q}_j^s+\tau_j^s({\bf x}) \dot{q}_j^s + \sigma_j^s({\bf x}) q_j^s\right] \nonumber\\
  &=&O\left(q_k^{s'}q_l^{s''},q_k^{s'}\dot{q}_l^{s''},\dot{q}_k^{s'}\dot{q}_l^{s''},q_k^{s'}\ddot{q}_l^{s''}\right)\,.
\end{eqnarray}
Here $\rho_j^s$, $\tau_j^s$, and $\sigma_j^s$ are tensor functions of the
spatial coordinates, and they depend on the background metric as well
as the corresponding wave function of the quasinormal mode.  The right
hand side of the equation has a complicated $\bf x$-dependence that we
have suppressed.


We would now like to project Eq.~\eqref{eq:pluggedin} onto individual
modes to obtain equations for a set of nonlinearly coupled oscillators
in the form of
\begin{eqnarray}\label{eq:projectedEinstein}
  &&a_j^s \ddot{q}_j^s+b_j^s \dot{q}_j^s + c_j^s q_j^s \nonumber\\
  &=&\hat{S}_j^s\left(q_k^{s'}q_l^{s^{\prime\prime}},q_k^{s'}\dot{q}_l^{s^{\prime\prime}},\dot{q}_k^{s'}\dot{q}_l^{s^{\prime\prime}},q_k^{s'}\ddot{q}_l^{s^{\prime\prime}}\right)\,,
\end{eqnarray}
for each $j$ and $s$.  In order to do so we require a suitable set of
projectors.  If, along any of the dimensions transverse to the radial
direction, the background metric possesses a suitable isometry group
so that this part of the wave function is described by tensor
harmonics (Fourier modes, tensor spherical harmonics, etc.) then it is
easy to project out this part by using an inner product.  The
remaining part (generally including the radial direction) is however
more problematic.

It is often the case that the equations can be written in the form of
a standard eigenvalue problem, $\ddot\Psi=-A\Psi$.  For normal modes,
one can define an inner product $\langle\chi|\eta\rangle$ with respect
to which $A$ is self-adjoint, and the modes are orthogonal.  One can
then use this inner product to define the projector.  For dissipative
systems with quasinormal modes, the eigenvalues are complex and $A$
cannot be self-adjoint.  Another problem is that often
$|\mathcal{Z}_{\mu\nu}^{(j\pm)}|\to\infty$ at the dissipative
boundaries of the system.  Nevertheless, it is still possible to
define a suitable bilinear form, with respect to which $A$ is
symmetric~\cite{Leung:1994,Leung:1998,Leung:1999rh,Yang:2014tla,Zimmerman2014pr,Yang:2014zva,Mark2014}.
This bilinear form involves an integral of $\chi\eta$ without any
complex conjugation so symmetry of $A$ does not imply that the
eigenvalues are real.  Furthermore it is still necessary to
appropriately regulate the integration to eliminate divergences.  The
bilinear form may be regarded as a ``generalized'' inner product, and
be used as such. In particular, it may then be shown that
$\langle\mathcal{Z}^{j\pm}|\mathcal{Z}^{k\pm}\rangle = 0$ for
$\omega_j\ne\omega_k$, and this orthogonality leads to a suitable
projector.

In the general case (such as the coordinate system we use in
Sec.~\ref{sec4}) it is not necessarily possible to re-write the
equation as a standard eigenvalue problem.  Nevertheless, we can still
define a generalized inner product and use it to project the equation
onto modes.  It may be that the modes are not orthogonal with respect
to this inner product, in which case the projection of the left hand
side of \eqref{eq:htruncated} contains contributions from additional
modes beyond the desired projection mode.  After performing
projections onto all modes, it would then be necessary to diagonalize
the system to obtain a set of equations of the
form~\eqref{eq:projectedEinstein}. This is possible by applying procedures described in Sec. \ref{sec:nondeg} to remove  ``unphysical modes" and reduce the order of the differential equations.
At this point, it is worth noting that in principle any inner product
which leaves this set of equations non-degenerate fits our
purpose. However, in order to minimize the error from neglecting the
residual part, it is good practice to adopt an inner-product suitable
for eigenvalue perturbation analysis (see Sec.~\ref{sec:innerprod} for a concrete example of such an inner-product).

With the equations decoupled as in~\eqref{eq:projectedEinstein} with a
suitable generalized inner product, we can now substitute in
Eq.~\eqref{eqqab} for $q_j^\pm$.  We obtain,
\begin{eqnarray}
  \label{eq:Addot}a_j^-\ddot{A}_j+\tilde{b}_j^-\dot{A}_j &=& S_j^-\left(A_k,B_l\right),\\
  \label{eq:Bddot}a_j^+\ddot{B}_j+\tilde{b}_j^+\dot{B}_j &=& S_j^+\left(A_k,B_l\right),
\end{eqnarray}
where $\tilde{b}_j^- \equiv b_j^--2i\omega_ja_j^-$ and
$\tilde{b}_j^+ \equiv b_j^++2i\omega_j^\ast a_j^+$.  We have used the
fact that $e^{-i\omega_j t}$ and $e^{i\omega_j^\ast t}$ are
homogeneous solutions to simplify the left hand sides.  The ``source''
terms on the right hand sides are quadratic in $A_k$ and $B_k$.  We
have dropped quadratic terms involving derivatives of $A_k$ and $B_k$
in $S_j^s$ as we expect them to be smaller than quadratic terms not
involving derivatives.  Indeed Eqs.~\eqref{eq:Addot}--\eqref{eq:Bddot}
already indicate that time derivatives of the coefficients are of
quadratic order in the perturbation amplitudes, so that, e.g., terms
on the right hand side of the form $A_k\dot{A}_l$ would be of cubic
order. In general, the nonlinear terms will then be of the form
\begin{eqnarray}
  S_j^-&=&\sum_{lk} \left[\kappa^{-(1)}_{jkl} A_k A_l e^{-i (\omega_k+\omega_l-\omega_j) t}+ {\kappa}^{-(2)}_{jkl} A_k B_l e^{-i (\omega_k-\omega^*_l-\omega_j) t}\right. \nonumber \\
&&\quad \left.+  {\kappa}^{-(3)}_{jkl} B_k B_l e^{i (\omega^*_k+\omega^*_l+\omega_j) t}\right],
\end{eqnarray}
where the coefficients $\kappa_{jkl}^{-(n)}$ are constants (and
similarly for $S_j^+$).

We now proceed to separately analyze non-degenerate and degenerate
modes.

\subsubsection{Non-degenerate modes}\label{sec:nondeg}

The non-degenerate case applies to quasinormal modes only.  We
immediately see from examining \eqref{eq:Addot}--\eqref{eq:Bddot} that
with $S_j^s=0$, $\{A_j,B_j\}=\text{ constants}$ are solutions.  This
is by design as (\ref{eqqab-const}) are solutions to the linearized
equations.  However, if $a_j^s\ne0$ the left hand sides of
\eqref{eq:Addot}--\eqref{eq:Bddot} are second order in time, so that
there are additional homogeneous solutions,
\begin{equation}
  A_j \propto e^{-\tilde{b}_j^+t/a_j^+},\quad B_j \propto e^{-\tilde{b}_j^-t/a_j^-},
\end{equation}
which give rise to
\begin{equation}
  q_j^+\propto e^{(i\omega_j-b_j^+/a_j^+)t},\quad q_j^- \propto e^{(-i\omega_j^\ast-b_j^-/a_j^-)t}.
\end{equation}
These solutions are clearly not quasinormal modes since when combined
with the spatial wavefunctions, they do not satisfy the appropriate
dissipative boundary conditions.  In addition, if we multiply them
with the wave function $\mathcal{Z}^{\pm}_j$, the original linearized
Einstein equation is  not necessarily satisfied (if $a^s_j \neq 0$ and
$b^s_j \neq 0$). At the
linear level, one can require $A_j, B_j$ to be constants to remove
these spurious modes. At the nonlinear level, we need a systematic
strategy to eliminate this extra unphysical degree of freedom. 


Let us first assume that $a_j^s\ne0$.  For clarity we only consider
the $s=+$ modes, but the analysis carries over directly to $s=-$.  We
will argue that the second time derivative terms in
equations~\eqref{eq:Addot} and \eqref{eq:Bddot} should be
dropped. To arrive at an intuition for this, first note that we are
considering the problem of mode excitation in the presence of
sources. In equations~\eqref{eq:Addot} and \eqref{eq:Bddot}, the
source terms come from nonlinear couplings, but it is more instructive
to move beyond this particular specialization and consider generic
sources.  If a delta-function source $S=\delta^{(4)}(x^\mu-x^\mu_0)$ is
introduced to the spacetime, it gives rise to a finite-value
discontinuity of the quasinormal mode amplitude at $t=t_0$, after
which quasinormal modes evolve freely and $A_j$ remains constant (see
the example in Appendix \ref{sec:schwarz}).  In other words, only
$A_j$ jumps at the delta source while $\dot{A}_j$ is unaffected 
(otherwise it will not remain constant in the ensuing free-evolution),
so that only $\dot{A}_j$ is needed in a sourced mode evolution
equation to account for the influence of that source, while $\ddot{A}$
does not in fact contribute to the evolution of the physical
modes. Furthermore, dropping $\ddot{A}$ also frees us of the
unphysical spurious modes, as the evolution equation is now first
order in time. We have subsequently
\begin{equation}\label{eq:ndg}
  \dot{A}_j=\frac{S_j^-}{\tilde{b}_j^+},\qquad\dot{B}_j=\frac{S_j^+}{\tilde{b}_j^+}.
\end{equation}

Mathematically, this physical intuition is reflected in the fact that
when we integrate \eqref{eq:Addot} from $t_{0-}$ to $t_{0+}$ with a
delta-function source at $t=t_0$, we realize that the integration of
the $\ddot{A}$ term in fact vanishes because $\dot{A}^s_j(t_{0-})$ and
$\dot{A}^s_j(t_{0+})$ must both be zero in order to satisfy the free
evolution condition when the source vanishes.  We note of course that
the solutions of equation~\eqref{eq:ndg} no longer strictly satisfy
the original equations \eqref{eq:Addot} or \eqref{eq:Bddot}. However,
since both set of equations should be satisfied on physical grounds,
$\ddot{A}$ and $\ddot{B}$ terms should be balanced by the residual
part of the metric perturbations, which is implicit in the left hand
sides of \eqref{eq:Addot} and \eqref{eq:Bddot}.

The situation with $a_j^s=0$ does not present any of the above
difficulties as the oscillator equation (\ref{eq:Addot}) or
(\ref{eq:Bddot}) is already first order in time, so that
\begin{equation}\label{eq:nondegena0}
  \dot{A}_j=\frac{S_j^-}{b_j^+},\quad\text{or}\quad\dot{B}_j=\frac{S_j^+}{b_j^+}.
\end{equation}
In fact, this is the case we shall encounter in Sec.~\ref{sec4} when
we perturb about the anti--de Sitter black brane background in ingoing
Eddington-Finkelstein coordinates.  In that case perturbations are
described by a first order in time and second order in space partial
differential equation.

\subsubsection{Degenerate modes}

For a degenerate mode, the two equations
in~(\ref{eq:projectedEinstein}) for $s=\pm$ degenerate to a single
equation for $q_j=q_j^-+q_j^+$.  Thus the 4 degrees of freedom present
for a given $j$ that we saw in the non-degenerate case reduce to 2
degrees of freedom (or 1 if $a_j=0$). In other words, we do not have
any unphysical spurious solutions in the degenerate case, but instead
two sets of physical solutions with the same spatial wavefunction, which
should both be kept. The consequence of this observation is that in
the end, the evolution equation for each mode is of first order, and
we need not apply the treatment for the $\ddot{A}$ term employed in the
non-degenerate case.
 
Consider first the case where $a_j\ne 0$.  As noted earlier, this
corresponds to a non-dissipative (i.e. normal) mode.  An example where this
occurs is in perturbations about pure anti--de Sitter spacetime
(without any black hole).  (The case of coupled scalar field-general
relativity perturbations about AdS was analyzed as coupled oscillators
within the context of a two timescale expansion
in~\cite{Balasubramanian:2014cja}.)

As discussed before, even for this $a_j\ne 0$ case, the
\eqref{eq:Addot}--\eqref{eq:Bddot} should reduce to first order, and
we show below how this is to be achieved.  First note that we have
\begin{equation} \label{eq:qvsAB2}
q_j = A_j(t) e^{-i\omega_j t} + B_j(t) e^{i\omega^*_j t}\,,
\end{equation}
and when we introduced time dependence into $A_j$ and $B_j$, these
parameters can in themselves contain $e^{-i\omega_j t}$ and
$e^{i\omega^*_j t}$ factors, so their choices in equation
\eqref{eq:qvsAB2} are not unique, and we have in effect a 
freedom that we have to fix. The most obvious optimal choice is to enforce
\begin{equation}
  \dot{q}_j=-i\omega_jA_je^{-i\omega_jt}+i\omega_j^\ast B_j e^{-\omega_j^\ast t}.
\end{equation}
as a gauge fixing, or equivalently 
\begin{equation}\label{eqqdab}
  \dot{A}_je^{-i\omega_jt}+\dot{B}_je^{i\omega_j^\ast t}=0,
\end{equation}
which incidentally looks as if we were solving an inhomogeneous equation through a variation of
parameters method. 
The physical intuition behind this constraint is that $A_j$ and $B_j$
change only slowly with time so it is appropriate to regard them as
``instantaneous'' amplitudes. (However, this does not constitute a restriction on the solution.)
We then have 
\begin{align}\label{eqab}
& A_j =e^{i \omega_j t}  \frac{\omega^* q_j+i \dot{q}_j}{\omega+\omega^*}\,,\nonumber \\
& B_j =e^{-i \omega^*_j t}  \frac{\omega^* q_j-i \dot{q}_j}{\omega+\omega^*}\,.
\end{align}
So far we have not imposed any equation of motion, and after
substituting in equation~\eqref{eq:projectedEinstein} and walking
through the same procedure as that presented in Appendix
\ref{sec:Oscillators},  we obtain
\begin{equation}\label{eqwant}
\dot A_j = \frac{i e^{i \omega_j t}}{ a_j (\omega_j^*+\omega_j)}  \hat{S}_j\,, \quad
\dot B_j = -\frac{i e^{-i \omega^*_j t}}{ a_j (\omega_j+\omega^*_j)}  \hat{S}_j\,. 
\end{equation}
We have thus re-expressed the second order
equation~(\ref{eq:projectedEinstein}) for $q_j$ in terms of first
order equations for the amplitudes $A_j$ and $B_j$ .

In the case where $a_j=0$, we have $\omega_j=-\omega_j^\ast=-
ic_j/b_j$, so $\omega_j$ is purely imaginary and there is a single
degree of freedom.  There is then no need to distinguish $A_j$ and
$B_j$, so we can set $B_j=0$. Equation~(\ref{eq:Addot}) easily reduces
to
\begin{equation}\label{eqwant2}
\dot A_j = \frac{S_j}{b_j} =\frac{\hat{S}_j e^{i \omega_j t}}{b_j}\,.
\end{equation}

Equations~\eqref{eq:ndg}, \eqref{eq:nondegena0}, \eqref{eqwant} and
\eqref{eqwant2} are our desired first order equations of motion. They
describe a collection of nonlinearly coupled harmonic oscillators. For any
suitable background spacetime, perturbations are characterized by the
mode spectrum, the mode-mode coupling coefficients and the mode
excitation factors.

Despite being a simplified model in the small amplitude limit, the
formalism we introduced in this section effectively serves as a
general platform to quantitatively compare and study the nature of
nonlinear gravitational phenomena in different spacetimes. A most
attractive feature is that the vast literature on nonlinear coupled
oscillators that has been developed in other branches of physics can
now be applied directly to the study of gravitational
interactions. For example, a precursor to the present procedure led to
the discovery of the parametric instability in the wave generation
process in near-extremal Kerr spacetimes in Ref.~\cite{Yang:2014tla},
which exhibited similar properties to the parametric instability in
nonlinear driven oscillators. In general relativity, another example is
furnished by the study of perturbed anti--de Sitter spacetimes through
a two timescale analysis~\cite{Balasubramanian:2014cja} and its
connection to the Fermi-Pasta-Ulam
problem~\cite{fpubook,2005Chaos..15a5104B}.

In Sec.~\ref{sec4} below (with some details relegated to Appendix
\ref{sec:Fluid}), we provide a concrete example on how to implement
the abstract procedure laid out in this section, using the
asymptotically AdS spacetime containing a black brane as the
background.  The study of this particular case also results in a
number of interesting physical observations, and so has its own
intrinsic value. For example, we shall see that relativistic
hydrodynamics admits a similar description to the gravitational
equations of motion, thus expanding the gravity/fluid
correspondence. Additionally, by connecting it to the fluid side one
concludes that the symmetry of $\kappa$ is closely connected to the
cascading/inverse-cascading behavior in the turbulent regime. Hence,
this duality mapping provides further evidence and insights for the
behavior of turbulence in gravity.

\section{\texorpdfstring{A\MakeLowercase{d}S}{AdS} black brane spacetimes and the gravity/fluid correspondence}\label{sec3}

In advance of our analysis of coupled AdS black brane quasinormal
modes in Sec.~\ref{sec4}, here we review the gravity/fluid
correspondence and study the black blane perturbations from the fluid
side.  We first present the background uniform AdS black brane solution. We then
review the derivative expansion method that leads to boundary fluid
equations that describe long wavelength perturbations.  Finally, by
Fourier transforming the boundary coordinates we re-write the system
as a set of coupled oscillators to facilitate comparison with our
later gravitational analysis.  For a more complete introduction to the
gravity/fluid correspondence, interested readers should consult the original
references~\cite{Baier:2007ix,Hubeny:2011hd,Bhattacharyya:2008jc,VanRaamsdonk:2008fp}.

\subsection{Background metric}

The metric for the $d+1$ dimensional uniformly boosted AdS black brane is given in ingoing Eddington-Finkelstein coordinates by
\begin{equation}\label{eq:metric0}
  ds^2_{[0]} = -2 u_\mu dx^\mu dr + 
 r^2 \left( \eta_{\mu\nu} + \frac{1}{(b r)^d} u_\mu u_\nu  \right) dx^\mu dx^\nu.
\end{equation}
where $u^\mu$ (with $u^\mu u_\mu = -1$) is some arbitrary constant
four velocity, $r$ is the radial coordinate and $x^\mu$ are the
boundary coordinates. The Hawking temperature of the black brane is
the constant $T =d/(4\pi b)$. This metric satisfies the Einstein
equation
\begin{equation}\label{eq:einsteineqns}
G_{\mu\nu}+\Lambda g_{\mu\nu}=0\,,
\end{equation}
with cosmological constant $\Lambda = -d(d-1)/2$.

Different choices of $u^\mu$ correspond simply to different
Lorentz-boosted boundary frames.  In particular, in the case where the
spatial velocity vanishes, the above metric simplifies to
\begin{equation}\label{eqingo}
ds^2_{[0]} = 2 dv dr -r^2 f(r) dv^2+ r^2\sum^{d-1}_{i=1} (dx_i)^2\,,
\end{equation}
where $f(r) \equiv 1-1/(b r)^d$ and $v=x^0$ is the ingoing Eddington-Finkelstein coordinate. The horizon is then located at $r=1/b$.  

If we define the tortoise coordinate $r_*$ as $d r_* = dr\, / (r^2 f(r))$ and $dv =dt +dr_*$, then the metric can be re-written
\begin{equation}
d s^2_{[0]} = r^2 \left [-f(r) dt^2+ \sum^{d-1}_{i=1} (dx_i)^2 \right ]+\frac{dr^2}{r^2 f(r)}\,,
\end{equation}
which is in the same form of Eq.~(4.1) of Ref.~\cite{Kovtun:2005ev}. Sometimes it is more convenient to work with a compactified radial coordinate, and normalize the boundary coordinates by the scale of the black brane horizon. With  $u \equiv 1/(b r)^2$, $\tilde{t} \equiv t\, 8\pi T/d $, $\tilde{x}^i \equiv x^i\,8 \pi T/d $ and $f(u) = 1-u^{d/2}$, the metric becomes
\begin{align}\label{eqkov}
d s^2_{[0]} = &\frac{( 4\pi T /d)^2}{ u} \left [ -f(u) dt^2+ \sum^{d-1}_{i=1} (dx^i)^2\right ]+\frac{1}{4 u^2 f(u)} du^2 \nonumber \\
=& \frac{1}{4 u} \left [ -f(u) d \tilde{t}^2+ \sum^{d-1}_{i=1} (d \tilde{x}^i)^2\right ]+\frac{1}{4 u^2 f(u)} du^2\,.
\end{align}

To derive the gravity/fluid correspondence, we take as our starting point the uniformly boosted black brane~\eqref{eq:metric0}.

\subsection{Gravity/fluid correspondence}

To each asymptotically AdS bulk solution there is an associated
metric and conserved stress-energy tensor on the timelike boundary of
the spacetime at $r\to\infty$ (see, e.g.,
Ref.~\cite{Balasubramanian:1999re}).  The boundary metric, in the case
of \eqref{eq:metric0} is $\eta_{\mu\nu}$, while the boundary
stress-energy tensor is
\begin{equation}
  T_{\mu\nu}^{[0]} = \frac{1}{16\pi G_{d+1} b^d}(du_\mu u_\nu + \eta_{\mu\nu}).
\end{equation}  
This describes a perfect fluid with energy density $\rho$ and pressure
$p$ given by
\begin{align}
  \rho &= \frac{d-1}{16\pi G_{d+1} b^d},\\
  p &= \frac{1}{16\pi G_{d+1} b^d}.
\end{align}
The stress-energy tensor is traceless, with equation of state
\begin{equation}
  p=\frac{\rho}{d-1},
\end{equation}
as required by conformal invariance.  Imposing the first law of
thermodynamics, $\mathrm{d}\rho = T \mathrm{d}s$, as well as the
relation $\rho+p = sT$, gives the entropy density $s$
and fluid temperature $T$,
\begin{align}
  s&=AT^{d-1},\\
  \rho&=\frac{d-1}{d}AT^d.
\end{align}
Here, $A$ is a constant of integration.  This is fixed to $A \equiv
(4\pi)^d/(16\pi G_{d+1} d^{d-1})$ by equating $T$ with the Hawking
temperature.

At this point, the fluid we have described is of constant density,
pressure and velocity.  To go beyond the uniform fluid, $b$ and
$u^\mu$ are promoted to functions of the boundary coordinates
$x^\mu$. Importantly, these will be assumed to vary slowly; that is,
if $L$ is the typical length scale of variation of these fields, then
$L\gg b$. With non-constant boundary fields, the metric
\eqref{eq:metric0} no longer describes a solution to the Einstein
equation. However, a solution can be obtained by systematically
correcting the metric order by order though a {\em derivative
  expansion}, so that the Einstein equation is solved to any desired
order in derivatives. One can then compute the boundary stress-energy
tensor corresponding to the metric at each order, and take this as
defining the boundary fluid.

After a rather long, but direct, calculation, the resulting
boundary stress-energy tensor (to second order in derivatives) is
\begin{equation}\label{eq:Tmunu2}
  T_{\mu\nu}^{[0+1+2]} = \frac{\rho}{d-1}\left(d u_\mu u_\nu + \eta_{\mu\nu}\right) + \Pi_{\mu\nu},
\end{equation}
where the viscous part $\Pi_{\mu\nu}$ is (see, e.g., Eq.~(3.11) of
Ref.~\cite{Baier:2007ix})
\begin{align}\label{eq:Pi1}
  \Pi_{\mu\nu}={}&- 2\eta \sigma_{\mu\nu} \nonumber\\
  & + 2\eta\tau_\Pi\left(\langle u^\alpha\partial_\alpha \sigma_{\mu\nu}\rangle + \frac{1}{d-1}\sigma_{\mu\nu}\partial_\alpha u^\alpha\right) \nonumber\\
 &+ \langle \lambda_1 \sigma_{\mu\alpha}\sigma_{\nu}^{\phantom{\nu}\alpha} + \lambda_2 \sigma_{\mu\alpha}\omega_{\nu}^{\phantom{\nu}\alpha} + \lambda_3 \omega_{\mu\alpha}\omega_{\nu}^{\phantom{\nu}\alpha}\rangle.
\end{align}
The shear and vorticity tensors are defined as,
\begin{align}
  \sigma_{\mu\nu} &\equiv \langle\partial_\mu u_\nu\rangle,\\
  \omega_{\mu\nu} &\equiv P_\mu^{\phantom{\mu}\alpha}P_\nu^{\phantom{\nu}\beta}\partial_{[\alpha}u_{\beta]}.
\end{align}
We have employed angled brackets to denote the symmetric traceless part of the projection
orthogonal to $u^\mu$,
\begin{equation}
  \langle A_{\mu\nu}\rangle \equiv \left(P_{(\mu}^{\phantom{(\mu}\alpha}P_{\nu)}^{\phantom{\nu)}\beta} - \frac{1}{d-1}P_{\mu\nu}P^{\alpha\beta}\right)A_{\alpha\beta},
\end{equation}
and defined $P_{\mu\nu}$ to be the spatial projector orthogonal to $u^\mu$,
\begin{equation}
  P_{\mu\nu} \equiv \eta_{\mu\nu} + u_\mu u_\nu.
\end{equation}
Notice that $\Pi_{\mu\nu}$ is symmetric and satisfies
\begin{align}
  \Pi^\mu_{\phantom{\mu}\mu}&=0,\\
  u^\nu\Pi_{\mu\nu}&=0.
\end{align}
The transport coefficients $\{\eta,\,\tau_\Pi,\,\lambda_i\}$ for
various dimensions can be found in, e.g.,
\cite{VanRaamsdonk:2008fp,Haack:2008cp,Bhattacharyya:2008mz}.  In
particular, $\eta=s/(4\pi)$. 

Projection of the Einstein equation along the boundary directions
shows that the boundary stress-energy tensor is conserved,
giving rise to the fluid equations of motion,
\begin{eqnarray}
0&=& u^\nu \partial_\nu \rho+\frac{d}{d-1} \rho \partial_{\nu} u^\nu - u^\mu \partial^\nu \Pi_{\mu\nu}\,, \\
0&=& \frac{d}{d-1} \rho u^\mu \partial_\mu u^\alpha +\frac{\partial^\alpha \rho}{d-1} -\frac{d}{(d-1)^2} u^\alpha \rho \partial_\mu u^\mu \nonumber \\
&&+\frac{1}{d-1} u^\alpha u^\mu\partial^\nu \Pi_{\mu\nu} +P^{\alpha\mu} \partial^\nu \Pi_{\mu\nu}\,.
\end{eqnarray}

\begin{figure}[t,b]
\includegraphics[width=0.90\columnwidth]{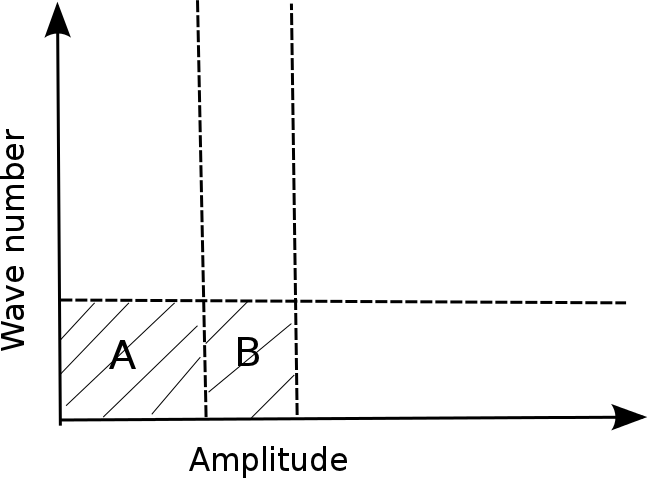}
\caption{An illustration of the hydrodynamical expansion (small wave
  number) and black-hole perturbation (small amplitude). They both
  admit effective coupled oscillator descriptions. In AdS black-brane
  spacetime we compare the results from both sides of the
  duality, in the shaded region of the plot.  For small perturbation amplitude, this comparison has been done in the linearized perturbation theory (as depicted by region ``A" and see for example \cite{Kovtun:2005ev}). For larger perturbation amplitude (region ``B"), we are able to expand the comparison to equations of motion \emph{with nonlinear couplings} using the coupled-oscillator model. }
\label{fig:comparison}
\end{figure}

The gravity/fluid correspondence thus provides an explicit link between
black hole perturbations in the sufficiently {\em long wavelength
  regime}---described by small wave numbers---and relativistic
hydrodynamics.  Ordinary perturbation theory, by contrast, provides a
solution that is valid for sufficiently {\em small amplitudes}, but
cannot easily capture the transfer of energy between modes.  Our
coupled-oscillator approach in contrast {\em does}
capture the leading mode-mode couplings that are manifest in the fluid
picture, and it is in that sense valid for larger amplitudes (see
Sec.~\ref{sec:larger}).  As illustrated in Fig.~\ref{fig:comparison},
there is an overlapping regime where the predictions of both
approaches can be compared.

\subsection{Mode expansion of the boundary fluid}

We now proceed to re-write the fluid equations as a set of coupled
oscillator equations so that they can be compared with the equations
we will derive on the gravity side.  We denote the four velocity
$u^\mu=(\gamma, {\bf u})$, where $\gamma^2=1+{\bf u} \cdot {\bf u}$,
and the density $\rho =\rho_0 e^\xi$.  Keeping viscous terms to linear
order in $\bf u$ and $\xi$, and inviscid terms to quadratic order (as
needed for the comparison), the energy conservation and Euler
equations reduce to
\begin{eqnarray}\label{eqmexpl}
0&=&\partial_t \xi +{\bf u} \cdot \nabla \xi+\frac{d}{d-1}(\partial_t \gamma+\nabla \cdot {\bf u})\,, \\
\label{eq:eulerapprox}0&=&\partial_t {\bf u} +{\bf u} \cdot \nabla {\bf u}+\frac{1}{d}\nabla \xi -\frac{1}{d-1} (\partial_t \gamma+\nabla \cdot {\bf u}) {\bf u}\nonumber\\
&&-\frac{\eta}{\rho_0}\left(\frac{d-1}{d}\nabla^2{\bf u}+\frac{d-3}{d}\nabla(\nabla\cdot{\bf u})\right)\,.
\end{eqnarray}
Furthermore, dropping nonlinear terms,
\begin{eqnarray}
0&=& \partial_t \xi^{(1)} +\frac{d}{d-1}\nabla \cdot {\bf u}^{(1)}\,, \label{eq:LinFluidEq1} \\
0&=& \partial_t {\bf u}^{(1)} +\frac{1}{d}\nabla \xi^{(1)} \nonumber\\
&&-\frac{\eta}{\rho_0}\left(\frac{d-1}{d}\nabla^2{\bf u}^{(1)}+\frac{d-3}{d}\nabla(\nabla\cdot{\bf u}^{(1)})\right) \label{eq:LinFluidEq2} \,.
\end{eqnarray}

Linearized solutions are decomposed into two families of modes: sound
and shear. A sound wave of momentum $\bf k$ takes the form
\begin{equation}
{\bf u}^{(1)}_b \sim A_b({\bf k}) e^{-i \omega_b t} e^{i {\bf k} \cdot {\bf x}} \hat {\bf k}\,, \quad \xi^{(1)} \sim B_b({\bf k}) e^{-i\omega_b t} e^{i {\bf k} \cdot {\bf x}}\,. 
\end{equation}
By solving the linearized equations \eqref{eq:LinFluidEq1} and
\eqref{eq:LinFluidEq2}, the dispersion relation is found to be
\begin{equation}
  \omega_b = \pm \frac{k}{\sqrt{d-1}} - i\frac{d-2}{d}\frac{\eta}{\rho_0}k^2+O(k^3)\,,
\end{equation}
and
\begin{equation}
  B_b({\bf k})=\frac{d}{d-1}\frac{k}{\omega_b}A_b({\bf k})\,.
\end{equation}
For the shear modes, $\xi^{(1)}=0$ and
\begin{equation}
  {\bf u}^{(1)}_s \sim A_s({\bf k}) e^{-i \omega_s t} e^{i {\bf k} \cdot {\bf x}} \hat {\bf u}_s\,,
\end{equation}
with $\hat {\bf u}_s \cdot {\bf k}=0$. The resulting dispersion
relation is
\begin{equation}\label{eq:shearfreqfluid}
  \omega_s=-i\frac{d-1}{d}\frac{\eta}{\rho_0}k^2+O(k^3)\,,
\end{equation}
so shear modes are purely decaying.  The general solution to the
linearized fluid equations is simply a sum over sound and shear modes
of different $\bf k$ and shear polarizations $s$.

We are now in a position to include the effects of nonlinear coupling
terms.  To do so, we express $\xi$ and $\bf u$ as sums over linear
modes, but we allow for the coefficients $A$ and $B$ to be
functions of time.  The velocity ansatz then takes the form
\begin{equation}\label{eqmexpan}
{\bf u}({\bf x}, t) = \sum_{\bf k} \left[q_b({\bf k}, t)   \hat {\bf k}+\sum_s  q_s({\bf k}, t) \hat {\bf u}_s\right] e^{i {\bf k} \cdot {\bf x}}\,,
\end{equation}
where $q_s({\bf k},t) = A_s({\bf k},t) e^{-i \omega_s t}$ and
$q_b({\bf k},t) = A_b({\bf k},t) e^{-i \omega_b t}$.  The coefficients
are of course subject to a reality condition.  Inserting this
expansion into Eq.~\eqref{eq:eulerapprox}, and projecting it onto a
particular shear mode, we obtain
\begin{eqnarray} \label{eqshear}
&&\partial_t A_s({\bf k}, t)   \\
&=&i \sum_{{\bf p}+{\bf q}={\bf k},\, s',\,{s''}} [\hat {\bf u}_{s'}({\bf p},t) \cdot {\bf q}][\hat {\bf u}_s ({\bf k},t)\cdot \hat {\bf u}_{s''}({\bf q},t) ] A_{s'}({\bf p},t) A_{s''}({\bf q},t) \nonumber \\
&&+\sum_{{\bf p}+{\bf q}={\bf k},\, s'} (\cdots) A_{s'}({\bf p},t) q_{b}({\bf q},t) 
+\sum_{{\bf p}+{\bf q}={\bf k}} (\cdots) q_{b}({\bf p},t) q_{b}({\bf q},t) \,,\nonumber
\end{eqnarray}
Notice that the left hand side has been reduced to simply the time
derivative of $A_s$ because the mode function satisfies the linearized
equation of motion.  The right hand side describes the nonlinear
coupling between modes.

The second and third terms (coupling coefficients unspecified) in
Eq.~\eqref{eqshear} describe the mixing between the sound modes and
the shear modes, as well as between two sound modes. The coefficients
to these terms contain fast [$\exp(i\omega t)$ type] oscillatory
time-dependent factors, so their effects tend to average to zero
during the longer time scales in which we examine the growth and decay
of modes. On the other hand, the first term describes the mixing
between two shear modes, and it trivially satisfies the ``resonant
condition'' in the time-domain since $\Re(\omega_s)=0$.  This results
in significant energy transfer between shear modes (and had we been
performing an ordinary perturbative expansion would have resulted in
secular growth).  It is then natural to expect that the effect of
sound modes is sub-dominant in the turbulent process of conformal
fluids, where the viscous damping is less important. In fact, if we
ignore all the sound modes in the relativistic hydro equation, the
resulting Eq.~(\ref{eqshear}) is the same as the one for
incompressible fluid (Appendix \ref{sec:Fluid}), and they share the same
conservation laws in the Fourier domain.

Equation~\eqref{eqshear} expresses the fluid as a collection of
coupled oscillators, to be compared with~\eqref{eqwant} on the gravity
side. In the next section we shall apply the general formalism of
Sec.~\ref{sec2} to the AdS black brane spacetime and directly match
its mode coupling coefficients (for the fundamental hydro shear
quasinormal modes) to the shear-shear mode coupling coefficients in
Eq.~\eqref{eqshear}. One can apply the same procedure to verify the
correspondence in the sound channel (which we have not written
down). We will only address the shear modes, as the main purpose of
this work is to formulate the coupled oscillator model and to
illustrate its technical details, rather than to provide a full
verification of the gravity/fluid correspondence. We envisage that
this framework shall prove its unique value when studying
gravitational interactions in spacetime without a clear gravity/fluid
correspondence, or in cases where the hydrodynamical (long-wavelength)
approximation becomes too restrictive.

\section{Linear and nonlinear gravitational perturbations of the
  \texorpdfstring{A\MakeLowercase{d}S}{AdS}$_5$
  black-brane}\label{sec4}

In this section we study gravitational perturbations about an
asymptotically AdS black brane within the context of the coupled
oscillator model.  We adopt this particular example for two reasons:
On the one hand, the boundary metric of the background spacetime is
flat, which simplifies calculations when performing wave function
projections. On the other hand, the gravity/fluid correspondence is
well established in this spacetime, and this allows us to compare
results obtained in the gravity and dual fluid pictures, as depicted
in Fig.~\ref{fig:comparison}. In particular, we shall focus on the
analysis of shear modes at both linear and nonlinear levels. We also
fix the spacetime dimension to $d+1=5$, although it is straightforward
to generalize the analysis below to other dimensions. For calculations
within this section, we make further simplifications by scaling the
coordinates such that $b=1$, so the horizon is located at $r=1$. This
means that we effectively choose $T=1/\pi$ so [see above
Eq.~\eqref{eqkov}]
\begin{align}\label{eq:CoordChoice}
x^i = \frac{1}{2} \tilde{x}^i\,, \quad k_i = 2 \tilde{k}_i\,.
\end{align}

\subsection{Linear perturbation}\label{sec:gravitylinear}

Linear quasinormal mode perturbations of AdS black branes have been
thoroughly analyzed in~\cite{Kovtun:2005ev}. There, the fundamental
(slowly decaying) quasinormal modes of the spacetime were shown to be
the same as the hydrodynamical modes of the boundary fluid. The
analysis was performed using the coordinate system of
Eq.~\eqref{eqkov}, whereas for our purposes it is more convenient to
use the ingoing coordinates of Eq.~\eqref{eqingo}. As discussed in
Appendix \ref{sec:TwoBasis}, choosing different coordinates leads to
different definitions for the modes. At the linear level there exists
a clean one-to-one mapping of modes in different bases as each
quasinormal mode is a solution to the linear Einstein
equation. However, when studying nonlinear perturbations, their
projection with respect to a mode-basis associated to a different
coordinate system leads to an expansion with a less direct
identification. In Appendix \ref{sec:TwoBasis} we illustrate this
point with a simple example describing a scalar field propagating on
Minkowski spacetime.

As demonstrated in \cite{Kovtun:2005ev}, linear perturbations of the
AdS black brane can be classified into shear, sound and scalar
sectors. In addition, as the boundary metric is flat, it is
straightforward to Fourier transform the metric components along the
boundary coordinates. The same logic applies when we adopt ingoing
coordinates. Without loss of generality, we consider a mode whose
boundary-coordinate dependence is $e^{i k z}$. For shear
perturbations, the relevant metric components are then
$h_{r\alpha}, h_{v\alpha}, h_{z\alpha}$, where $\alpha=x,y$.  Without
loss of generality, we choose the polarization $\alpha =x$, and impose
the radial gauge condition $h_{rM} =0$, with $M =
\{r,v,z,x,y\}$. Defining the auxiliary variables
\begin{align}
H_{zx} \equiv h_{zx} \frac{e^{- i k z}}{r^2}, \quad H_{vx} \equiv h_{vx} \frac{e^{-i k z}}{r^2} \,,
\end{align}
the independent components of the linearized Einstein equation take
the form
\begin{eqnarray}\label{eqshearcompo}
  0&=&5 r \frac{\partial H_{vx}}{\partial r}+i k \frac{\partial H_{zx}}{\partial r}+r^2\frac{\partial^2 H_{vx}}{\partial r^2}\,,\\
  0&=&k^2 H_{vx}-5 r^3 f \frac{\partial H_{vx}}{\partial r}-r^4 f \frac{\partial^2 H_{vx}}{\partial r^2}+i k\frac{\partial H_{zx}}{\partial v}-r^2\frac{\partial^2 H_{vx}}{\partial v\partial r}\,.\nonumber
\end{eqnarray}

We can further simplify this system by defining the master variable,
$\Psi \equiv \partial_r H_{vx}$.  This satisfies the master equation,
\begin{equation} \label{eq:Psieq}
-k^2 \Psi+(5 r^3 f \Psi)'+(r^4 f \Psi')'+7r \dot \Psi+2r^2 \dot \Psi'=0\,,
\end{equation}
where in this section we will often denote partial derivatives as
$(\cdot)' \equiv \partial_r$ and $\dot{(\cdot)} \equiv \partial_v$.
To look for quasinormal modes, we first take advantage of the time
translation symmetry of the equation to impose a $e^{-i\omega v}$ time
dependence (so $\dot\Psi \to -i\omega \Psi$).  Solving the remaining
spatial equation with appropriate boundary conditions at the horizon
and spatial infinity gives rise to a set of quasinormal modes in the
ingoing coordinates, and the frequency spectrum $\omega(k)$.

To analyze the horizon boundary, we multiply Eq.~\eqref{eq:Psieq} by
$f$ and take the horizon limit $r \rightarrow1$.  The wave equation
becomes
\begin{equation}
(\partial^2_{r_*}+2 \partial_v \partial_{r_*})\Psi=0\,,
\end{equation}
with two independent solutions,
\begin{equation}
\partial_{r_*} \Psi=0,\quad {\rm and} \quad ( \partial_{r_*} +2 \partial_v )\Psi=0\,.
\end{equation}
The ingoing boundary condition for the quasinormal modes selects
\begin{equation}
\frac{\partial \Psi}{\partial r_*} \to 0,\quad r \rightarrow 1\,.
\end{equation}
As $r\to\infty$ we impose a reflecting boundary condition (since the
spacetime is asymptotically AdS), so the metric perturbation is
required to vanish. This means that we should at least expect
$h = O(1/r)$ and $\Psi = O(1/r^4)$.

The above discussion applies to all quasinormal modes of our system.
However, the dual fluid captures only the longest lived shear and
sound modes, which have $\omega\to0$ as $k\to0$ (known as the
``hydro'' modes).  In order to compare our results with the fluid we
therefore restrict to $\tilde k\ll1$.  We can then construct the
eigenfunctions perturbatively in $k$ (and $\omega$).  In this
expansion, the leading order part of equation \eqref{eq:Psieq} is
\begin{equation}
(5 r^3 f)' \Psi+5 f r^3 \Psi'+r^4 f \Psi''+(r^4 f)'\Psi'=0\,.
\end{equation}
After imposing the horizon boundary condition, the solution is
\begin{equation}
\Psi_0 = \frac{C(v)}{r^5}\,.
\end{equation}
where the subscript $0$
indicates that this solves the leading order equation. (Notice
that this solution also falls off sufficiently rapidly at spatial
infinity.)  To look for quasinormal mode solutions we take $C(v)=e^{-i\omega v}$.

The leading order solution $\Psi_0$ then sources the first order
correction $\Psi_1$ through
\begin{align}
&(5 r^3 f)' \Psi_1+5 f r^3 \Psi_1'+r^4 f \Psi_1''+(r^4 f)'\Psi_1' \nonumber \\
&= -(-k^2 \Psi_0+7r \dot \Psi_0+2r^2 \dot \Psi'_0)\,.
\end{align}
The combined solution $\Psi=\Psi_0+\Psi_1$ is then
\begin{align}
\Psi = &\left[\frac{1}{r^5}+\frac{ -(k^2-4 i \omega ) \log (1-r) -(k^2+4 i \omega) \log (1+r) }{16 r^5}\right. \nonumber \\
 & \left.+\frac{8 i \omega \arctan r +k^2\log (1+r^2)}{16 r^5}\right]e^{-i\omega v}\,.
 \end{align}
 In order to satisfy the horizon boundary condition we must impose
 $k^2 = 4 i \omega$, resulting in
\begin{equation}
\Psi = \left[\frac{1}{r^5}+\frac{2 k^2 \arctan r  -2k^2\log (1+r) +k^2\log (1+r^2)}{16 r^5}\right]e^{-i\omega v}\,.
\end{equation}
Using Eq.~\eqref{eq:CoordChoice}, we verify that $k^2 = 4 i \omega$ is
equivalent to $\tilde \omega = -i \tilde k^2 /2$, which is exactly the
dispersion relation of shear hydro quasinormal modes derived in
\cite{Kovtun:2005ev} using a different coordinate system.   In addition, it is easy to check that the
dispersion relation matches~\eqref{eq:shearfreqfluid}, derived on the
fluid side. 

Knowing $\Psi$, it is straightforward to use Eq.~\eqref{eqshearcompo}
to reconstruct the metric perturbations. For the shear modes
considered here, the metric perturbation is
\begin{eqnarray}
 h_{vx}&=& -\frac{A}{4r^2} e^{-i \omega v+i k z}\left[1+\frac{k^2r^2}{16}(\pi r^2 - 4r +2)\right.\nonumber\\
 &&\quad\left.-\frac{k^2}{16}(r^4-1)\left(2\arctan r +\log\frac{1+r^2}{(1+r)^2}\right)\right]\,,\nonumber \\ 
 \\
 h_{zx} &=& i \frac{A}{4} k r^2 e^{-i \omega v+i k z}\left[\frac{\pi}{4}-\frac{1}{r}-\frac{\arctan r}{2}\right.  \nonumber\\
 &&\quad\left. +\frac{1}{4}\log \frac{(1+r^2)(1+r)^2}{r^4}\right]\,.
\end{eqnarray}

\subsection{Mode projection\label{sec:innerprod}}

Having carried out the linear analysis, we are almost ready to
calculate the shear-shear mode coupling coefficient. There is one more
problem to tackle however, which is to project the Einstein equation
onto an individual mode to see how a source term affects its
evolution.  As described in Sec.~\ref{sec:ModeExp}, we adopt a
technique that has been proven very powerful in solving similar
problems \cite{Leung:1999rh,Leung:1999iq,Yang:2014tla, Mark2014,
  Yang:2014zva, Zimmerman2014pr}. Namely, we enlist a suitable
bilinear form to project the equation onto individual modes.

For later convenience, we define $\phi = r^5 \Psi$, so that
Eq.~(\ref{eq:Psieq}) takes the form
\begin{equation} \label{eq:phiEqlinear}
\left( \frac{f}{r} \phi'\right)'-k^2 \frac{\phi}{r^5}+\frac{7}{r^4}  \dot \phi+2 r^2 \left(\frac{\dot \phi}{r^5}\right)'=0\,.
\end{equation}
Fourier transforming the wave operator in $v$, we define
\begin{equation}\label{eqphih}
H_\omega \phi \equiv \left( \frac{f}{r} \phi'\right)'-k^2 \frac{\phi}{r^5}-\frac{ 7i \omega }{r^4} \phi-2 i \omega  r^2  \left( \frac{\phi}{r^5}\right)'\,.
\end{equation}
We also define a generalized inner product,
\begin{equation} \label{eq:InnerProd}
\langle \chi | \eta \rangle = \int^\infty_1 dr  \chi \, \eta\,.
\end{equation}
The operator $H_\omega$ is not symmetric under this bilinear form,
i.e.,
$\langle \chi | H_{\omega} \eta \rangle \neq \langle H_{\omega} \chi |
\eta \rangle $,
because of the fourth term in $H_{\omega}$. However, in the
hydrodynamic limit ($\tilde{k} \ll 1$) this term is neglected, so
\eqref{eq:InnerProd} is suitable for our purpose of comparing to the
dual fluid.  

For completeness, we note that should the need arises for the study of perturbations of
higher overtones away from the hydro limit, we may use an alternative
bilinear form (dependent on $\omega$) with respect to which
$H_{\omega}$ is symmetric so that
$\langle \chi | H_{\omega} \eta \rangle_{\omega} = \langle H_{\omega}
\chi | \eta \rangle_{\omega}$. In this case,
\begin{eqnarray} \label{eq:InnerProd2}
\langle \chi | \eta \rangle_{\omega} &=& \int^\infty_1 dr  g_\omega(r ) \chi \, \eta\,,\quad\text{with}  \\
\log g_\omega(r )&=&-i \omega \left ( \arctan r+\frac{1}{2}\log \frac{1-r}{1+r}\right )+{\rm const}\,,\nonumber
\end{eqnarray}
is the unique option. There is one $g_{\omega}$ for each $\omega$, so
we have a family of such generalized inner products. Using
$g_{\omega}$ to project onto the mode with frequency $\omega$
[followed by a diagonalization procedure as per the discussion above
Eq.~\eqref{eq:Addot}] is  a natural choice, and indeed
leads to agreement with the Green's function method for projecting
modes (see Appendix \ref{sec:schwarz}). In any case, to $O(k^2)$,
these generalized inner products reduce to Eq.~\ref{eq:InnerProd}.

For the purpose of the time-domain analysis in the next section, we
expect the effect of non-hydrodynamical modes [see
Eq.~\eqref{eq:NonHy} below] and the excitation of residual
parts\footnote{\label{fn5}The prompt piece of the residual can be intuitively understood
  as the source terms propagating on the light-cone. Also notice that
  the source terms, as represented by Eq.~\eqref{eqshear} or
  Eq.~\eqref{eqkpq}, are linear in the hydrodynamical momentum, so
  overall the source terms are of $O(k)$, as is the excitation
  amount of the prompt residual.}  to be at least $O(k)$. Therefore only
the hydrodynamical modes are important to leading order and we shall
adopt the generalized inner product \eqref{eq:InnerProd} for
calculations, as it is easier to implement in the time-domain
analysis. As an example, we show below that this inner product
generates the correct leading order (in $k$) frequency in the
eigenvalue analysis.

Let us now consider a simple example that demonstrates the essence of
how to utilize this inner product to carry out perturbation
studies. Suppose we perturb $k$ to $k+\epsilon \delta k$
($\epsilon \ll 1$) and ask for the change of $\omega$. On the one
hand, based on the dispersion relation $\omega= -i k^2/4$, we
immediately know that $\delta \omega = -i k \delta k/2$. On the other
hand, we can arrive at the same conclusion through a perturbation
analysis of the eigenvalue problem defined by Eq.~\eqref{eqphih}.

The change $k \rightarrow k+\epsilon \delta k$ causes $H $ to pick up
an extra term, $-2 \epsilon k \delta k/r^5$. We expect both the
eigenfrequency and the eigenfunction to also change to order
$\epsilon$,
\begin{align}
& \phi \rightarrow \phi+\epsilon \phi^{(1)}+ O(\epsilon^2)  \,, \nonumber \\
& \omega \rightarrow \omega + \epsilon \delta \omega +O(\epsilon^2)\,.
\end{align}
Plugging into the wave equation Eq.~\eqref{eqphih}, and projecting
both sides onto $\phi$ while keeping only the $O(\epsilon)$ terms, we
can eliminate the unknown function $\phi^{(1)}$ to obtain
\begin{align}
- i \delta \omega = & 2 k \delta k \frac{\langle \phi | 1/r^5 \phi \rangle}{ \langle \phi | 7\phi/r^4 +2r^2(\phi/r^5)' \rangle}+O(k^2) \nonumber \\
\approx &- \frac{k}{2} \delta k\,,
\end{align}
which is consistent with our expectation. We note that it was necessary in this
analysis to use the symmetry property of $H_\omega$ to eliminate terms
involving $\phi^{(1)}$.  Although somewhat
excessive for this simple problem, we see that with the
help of our generalized inner product, it is now possible to carry out
a perturbation analysis in a manner analogous to the application of
perturbation theory in quantum mechanics~\cite{Shankar} (for a direct
mapping of a wave equation with outgoing boundary condition into a
Schr\"odinger equation with non-Hermitian Hamiltonian,
see~\cite{Leung:1998}).

\subsection{Nonlinear analysis}\label{sec43}

We are now in a position to move beyond the linear level and study the
second order (nonlinear) Einstein equation~(\ref{eq:htruncated}).  We
begin by considering its projection onto the shear sector with spatial
dependence $e^{ikx}$ and spatial polarization $\alpha=z$ (see
Sec.~\ref{sec:gravitylinear}).  (It is straightforward to perform this
projection onto a Fourier basis element with an ordinary inner
product.  The nontrivial aspect is the subsequent projection onto the hydro
mode.)  The non-vanishing $vx$ and $rx$ components of the Einstein
equation take the form
\begin{align}\label{eqrs}
& 5 r \frac{\partial H_{vz}}{\partial r}+i k \frac{\partial H_{zx}}{\partial r}+r^2\frac{\partial^2 H_{vz}}{\partial r^2}=\tau_{rz} \,,\nonumber \\
&k^2 H_{vz}-5 r^3 f \frac{\partial H_{vz}}{\partial r}-r^4 f \frac{\partial^2 H_{vz}}{\partial r^2}+i k\frac{\partial H_{zx}}{\partial v}-r^2\frac{\partial^2 H_{vz}}{\partial v\partial r} \nonumber \\
&=\tau_{vz}\,.
\end{align}
We have formally written the nonlinear terms as ``sources'' on
the right hand side of the equation. At
quadratic order the nonlinear terms are
\begin{equation}\label{eq:fourierprojection}
\tau_{rz} \equiv -\langle e^{- i k x}, 2 R^{(2)}_{rz} \rangle, \quad \tau_{vz}  \equiv -\langle e^{- i k x}, 2 R^{(2)}_{vz} \rangle\,.
\end{equation}
The inner product $\langle\cdot,\cdot\rangle$ is the ordinary inner
product over the boundary spatial coordinates.  Equation~\eqref{eqrs}
is simply~\eqref{eqshearcompo} with nonlinear terms included, and a
simple switch of coordinates $x \leftrightarrow z$.

\begin{figure}[t,b]
\includegraphics[width=0.66\columnwidth]{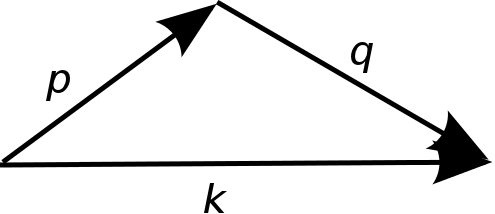}
\caption{An illustration of three wave numbers satisfying the ``momentum matching'' condition.} 
\label{fig:trig}
\end{figure}

Since the second order Ricci tensor is a quadratic function of
the metric perturbation, which can be expanded over Fourier
modes (and scalar, sound, shear sectors), the
projection~\eqref{eq:fourierprojection} enforces a wave number
matching condition on the terms that can contribute to the right hand
side of~\eqref{eqrs}.  Namely, modes with wave numbers ${\bf p}$ and
${\bf q}$ can only act as a source for mode ${\bf k}$ if
${\bf p}+{\bf q}={\bf k}$ (see Fig.~\ref{fig:trig}).  [This of course
also holds for the fluid analysis in~\eqref{eqshear}.]  We define the
angles $\theta_1 \equiv \arccos(\hat{q} \cdot \hat{k})$ and
$\theta_2 \equiv \arccos(\hat{p} \cdot \hat{k})$.

Following the same procedure as in the linear analysis, we re-write Eq.~\eqref{eqrs} in the form
of a sourced version of Eq.~\eqref{eq:phiEqlinear},
 \begin{eqnarray}\label{eqweqs}
&&\left(\frac{f}{r} \phi'\right)'-k^2 \frac{\phi}{r^5}+\frac{7}{r^4}  \dot \phi+2 r^2  \left( \frac{\dot \phi}{r^5}\right)' \nonumber \\
&=& - \dot \tau_{rz} -  \tau_{vz}' 
\equiv S_{\rm in}\,.
\end{eqnarray}
Since only first order time derivatives appear in this wave equation
and we know from the previous subsection that the quasinormal
frequency is purely imaginary, this shear hydrodynamic mode belongs to
the class described by Eq.~\eqref{eqwant2}.
 
 We now proceed to compute the nonlinear
source $S_j$ [see~\eqref{eqwant2}] using the generalized inner product
of Sec.~\ref{sec:innerprod}.  As in Sec.~\ref{sec:ModeExp}, we first
express the field $\phi$ as a sum over radial modes
\begin{equation}\label{eq:phiansatz}
 \phi \to A_0(t)e^{-i\omega_0 v} \chi_0(r )+\sum_{j>0} A_j(t) e^{-i \omega_j v} \chi_j(r ) +O(k)\,,  
\end{equation}
but we allow for the modes to have additional time dependence through
the mode amplitudes $A_j$.  Here the spatial wavefunctions are denoted
$\chi_j$, with $j=0$ corresponding to the hydro mode.  The non-hydro
modes all have frequencies $\omega_j=O(1)$, while $\omega_0=-ik^2/4$.
While $\phi$ is to be matched to modes of the $4$-velocity $u^\mu$ on the fluid side, we normalize the wave function
\begin{equation}
  \chi_0=4
\end{equation}
accordingly \footnote{This is of
  course just an inconsequential overall constant rescaling of $A$,
  the more important goal is to match the angular dependence of the
  coupling constants.}.  The $O(k)$
appearing in the expression for $\phi$ includes the residual
contribution under the hydrodynamical approximation (see Footnote
\ref{fn5}).


We can now plug \eqref{eq:phiansatz} into the wave
equation~\eqref{eqweqs}, and then take the generalized inner product
of both sides with $\chi_0$ using~\eqref{eq:InnerProd}.  Within this
computation, the effect of the non-hydrodynamical terms is at least
$O(k)$ [in fact $O(k^2)$] as we claimed in
Sec.~\ref{sec:innerprod}. This is because $e^{-i\omega_j v}\chi_j$
solves the linear equation~\eqref{eq:phiEqlinear}, so for $j>0$
\begin{eqnarray} \label{eq:NonHy}
  && \omega_j \left\langle \chi_0 \left| \frac{7}{r^4}\chi_j +2r^2\left(\frac{\chi_j}{r^5}\right.\right)' \right\rangle \nonumber \\
  &=& - \left\langle \chi_0 \left| \left( \frac{f}{r} \chi_j'\right )'-k^2 \frac{\chi_j}{r^5} \right.\right\rangle \nonumber \\
  &=& -\left\langle \left( \left. \frac{f}{r} \chi_0' \right)' - k^2 \frac{\chi_0}{r^5} \right| \chi_j \right\rangle \nonumber \\
  &=& \omega_0 \left\langle \left. \frac{7}{r^4}\chi_0 +2r^2\left(\frac{\chi_0}{r^5}\right)' \right| \chi_j \right\rangle \nonumber \\
  &=& O(k^2)\,.
\end{eqnarray}
Given this observation, it is now simple to show that 
\begin{eqnarray}\label{eqdrive}
  \dot A &\approx& \frac{1}{4} \frac{\langle \chi_0 | S_{\rm in} \rangle e^{i\omega_0v}}{ \langle \chi_0 | 7\chi_0/r^4 +2r^2(\chi_0/r^5)' \rangle} \nonumber \\
         &\approx& -\frac{1}{4} \tau_{vz} |_{r=1} \,,
\end{eqnarray}
where we dropped high order [$O(k^2)$ and higher] terms in $k$,
including nonlinear terms containing time derivatives (as discussed in
Sec.~\ref{sec:ModeExp}).  Using
Eq.~\eqref{eqr2}, the mode expansion of $h_{\mu\nu}$, and after some
lengthy but nevertheless straightforward calculations, one can show
that the shear-shear mode coupling coefficient arising
from~\eqref{eqdrive} is
\begin{eqnarray}
\kappa_{kpq} &=&  i k \sin (\theta_2-\theta_1) \,,
\end{eqnarray}
which agrees with the result obtained with its fluid counterpart from Eq.~\eqref{eqshear}
\begin{equation}\label{eqkpq}
\kappa_{kpq} = i [\hat {\bf u}_{s'}({\bf p},t) \cdot {\bf q}][\hat {\bf u}_s ({\bf k},t)\cdot \hat {\bf u}_{s''}({\bf q},t) ] +({\bf p} \leftrightarrow {\bf q})\,.
\end{equation}

We end this section by noting that the agreement between the mode
coupling coefficients inferred from the fluid equations and the AdS
black brane perturbation theory relies on the fact that they are
computed using the same mode basis, and that the comparison is made in
the regime where $\tilde{k} \ll 1$ and $|h| \ll 1$
(cf. Fig.~\ref{fig:comparison}). However, the coupled oscillator model
is applicable more broadly.

\section{Conclusions}\label{conclusion}

The study of nonlinear wave phenomena is undoubtedly a fascinating subject.
Gaining understanding in the particular case of general relativity poses
unique challenges even given the fixed speed of propagation of
physical perturbations. These challenges are rooted in the covariant
nature of the theory and physical degrees of freedom often hidden
within a larger set of (metric) variables.  These issues have hampered
understanding of gravitational perturbations beyond linear order
except in a few specialized
regimes~\cite{Gleiser:1998rw,Ioka:2007ak,Brizuela:2009qd,Pound:2012nt,Gralla:2012db},
seamingly leaving full numerical simulations as the main tool to try
to understand these issues (for a recent overview of these efforts,
see~\cite{chotuiklehnerpretorius} and references cited therein).

In the current work, we have presented a model to capture the
nonlinear behavior of gravitational perturbations\footnote{In this
  work we have included up to three-mode interactions, but the
  formalism can be extended to include higher order
  interactions.}. This model regards the system as composed of a
collection of nonlinearly coupled (damped) harmonic oscillators with
characteristic (isolated) frequencies given by quasinormal modes. By
construction this model reproduces standard results obtained at the
linearized level. At the nonlinear level, it describes mode-mode
couplings and their effect on frequency and amplitude shifts.  As an
illustration, we have shown how our model reproduces recent results
captured through the gravity/fluid correspondence via a purely
gravitational calculation. Importantly, the applicability of our
formalism is not restricted to long-wavelength perturbations---as in
the case of the gravity/fluid correspondence---so the coupled
oscillator model can also treat so-called ``fast (non-hydrodynamical)
modes'' of perturbed black holes~\cite{Friess:2006kw}.  As a
consequence it can be employed to study a broder phenomenology than
that reachable via the correspondence\footnote{Recently, resummation
  techniques have been proposed to take some of these higher modes
  into account within an extended hydrodynamical
  description~\cite{Heller:2013fn}. This requires knowledge of the
  hydrodynamical expansion to very large orders.}.  We stress that our
formalism is also applicable beyond asymptotically AdS spacetimes.
Thus it can also help shed light on nonlinear mode generation in
perturbations of asymptotically flat black hole
spacetimes~\cite{Papadopoulos:2001zf,Zlochower:2003yh}.

\acknowledgements

We thank David Radice for stimulating discussions about turbulent
fluids, Vitor Cardoso for further insights into perturbations of AdS
spacetimes as well as Michal Heller and Olivier Sarbach for general
discussions.  This work was supported in part by NSERC through a
Discovery Grant and CIFAR (to LL). FZ would like to thank the
Perimeter Institute for hospitality during the closing stages of this
work. Research at Perimeter Institute is supported through Industry
Canada and by the Province of Ontario through the Ministry of Research
\& Innovation.

\newpage
\appendix 

\section{Brief overview of coupled oscillator systems} \label{sec:Oscillators}

Consider a family of nonlinearly coupled harmonic oscillators governed by, 
\begin{align}\label{eqeom}
&\ddot q_j +\gamma_j \dot q_j+\tilde \omega^2_j q_j  \nonumber \\
&=\sum_{kl} ( \tilde{\lambda}^{(1)}_{jkl} q_k q_l + \tilde{\lambda}^{(2)}_{jkl} \dot q_k q_l+  \tilde{\lambda}^{(3)}_{jkl} \dot q_k \dot q_l) \equiv S_j\,, 
\end{align}
where $\tilde \omega^2_j q_j$ is the restoring force and $\gamma_j$ is the damping coefficient. Each oscillator's displacement can be decomposed 
in the same way as Eq.~\eqref{eqqab}, with $\omega_j$ satisfying
\begin{equation}
-\omega^2_j - i \gamma_j \omega_j+\tilde \omega^2_j=0\,.
\end{equation}

In the presence of nonlinear mode-mode coupling ($\tilde{\lambda}^{(n)}_{jkl}\ne0$),  $A_j$ and $B_j$ are both time-dependent. In fact, we can take one more time derivative of the first equation in Eq. (\ref{eqab})
, and obtain
\begin{align}
&(\dot A_j - i \omega _j A_j ) e^{- i \omega_j t} = \frac{1}{\omega_j+\omega^*_j} (\omega^*_j \dot q_j+i \ddot q_j) \nonumber \\
& =\frac{i S_j}{\omega_j+ \omega^*_j} +\left (\frac{\dot q_j \omega_j^*}{\omega_j+\omega^*_j} +\frac{\gamma_j \dot q_j+\tilde \omega^2_j q_j }{i (\omega^*_j+\omega_j)} \right )\,, \nonumber \\
& = \frac{i S_j}{ \omega_j+ \omega^*_j} - \frac{i \omega_j}{\omega_j+\omega^*_j} (\omega^*_j q_j+i \dot q_j)\,,
\end{align}
such that
\begin{equation}
\dot A_j = \frac{i S_j}{ \omega_j+ \omega^*_j} e^{i \omega_j t}\,,
\end{equation}
and similarly
\begin{equation}
\dot B_j =- \frac{i S_j}{ \omega_j+ \omega^*_j} e^{- i \omega^*_j t}\,.
\end{equation}
These effective equations of motion have the same kind of first-order form as Eq.~\eqref{eqwant2} and Eq.~\eqref{eqwant}, which means that one can utilize results from previous studies on nonlinear coupled oscillators to analyze nonlinear gravitational interactions.

\section{Two-dimensional incompressible fluid in the inertial regime \label{sec:Fluid}}

Here we review the Navier-Stokes equation for a two-dimensional
incompressible fluid. This discussion highlights how a new symmetry
for the mode-mode coupling coefficient arises in the mode-expansion
picture. Such symmetry is critical for the double-cascading (inverse
energy and direct enstrophy cascades) behavior in two-dimensionalfluids. A more
detailed discussion can be found in Ref.~\cite{Kraichnan1967}.

The Navier-Stokes equation for an incompressible fluid in the spatial-frequency domain reads
 \begin{align}
 &\left (\frac{\partial}{\partial t}+\nu k^2 \right ) u_j({\bf k}, t) \nonumber \\
 &= i k_l P_{jn}({\bf k})\sum_{{\bf p}+{\bf q}={\bf k}} u_n({\bf p}, t) u_l({\bf q}, t)\, \nonumber \\
 &=\frac{i k_l P_{jn}({\bf k})}{2}\sum_{{\bf p}+{\bf q}={\bf k}} [u_n({\bf p}, t) u_l({\bf q}, t) + u_n({\bf q}, t) u_l({\bf p}, t)]\,
 \end{align}  
where ${\bf u}({\bf x}, t) = \sum_{{\bf k}} e^{i {\bf k} \cdot {\bf x}} {\bf u}({\bf k},t)$ and $P_{jn}({\bf k}) \equiv \delta_{jn} -k_j k_n/k^2$. In incompressible fluids, the condition $\nabla \cdot {\bf u}=0$ translates to ${\bf k} \cdot {\bf u}({\bf k}, t)=0$ in the Fourier domain. We can write ${\bf u}({\bf k}, t)$ as
\begin{equation}
{\bf u}({\bf k}, t) =A({\bf k},t) {\hat u}({\bf k}, t) \,,
\end{equation}
where $\hat u({\bf k},t )$ satisfies $\hat u\cdot {\bf k} =0$ and $\hat u\cdot \hat u=1$. In $2+1$ fluids, $\hat u$ is unique for any ${\bf k}$. Using the new variables, the Navier-Stokes equation can be rewritten as
\begin{widetext}
\begin{align}
 \left (\frac{\partial}{\partial t}+\nu k^2 \right ) A({\bf k}, t)  &= i \sum_{{\bf p}+{\bf q}={\bf k}} \kappa({\bf k},{\bf p},{\bf q})A({\bf p}, t) A({\bf q}, t)\, \nonumber \\
 &=i  \sum_{{\bf p}+{\bf q}={\bf k}}\{[\hat u({\bf k}, t) \cdot \hat u({\bf p},t) ] [{\bf k} \cdot \hat u({\bf q},t)]+[\hat u({\bf k},t) \cdot \hat u({\bf q},t)] [{\bf k} \cdot \hat u({\bf p},t)]\}A({\bf p}, t) A({\bf q}, t) \,.
 \end{align}  
This is the same as the shear-shear coupling term in Eq.~\eqref{eqshear}, which is already written in a form consistent with the coupled oscillator model.

In the inertial regime we shall set the viscosity coefficient $\nu$ to zero (as such coefficient only governs the extent of the
regime but not the behavior within it) and recall that ${\bf u}({\bf x}, t)$ must be real. One can then show that
\begin{align}
-\frac{\partial [u_j({\bf k}, t) u^*_j({\bf k}, t)]}{\partial t}= 
 \sum_{{\bf p}+{\bf q}+{\bf k}=0}{\rm Im}\{[{\bf u}({\bf k},t) \cdot {\bf u}({\bf p},t) ] [{\bf k} \cdot {\bf u}({\bf q},t)]+[{\bf u}({\bf k},t) \cdot {\bf u}({\bf q},t) ] [{\bf k} \cdot {\bf u}({\bf p},t)]\} \nonumber \\ \equiv \sum_{{\bf p}+{\bf q}+{\bf k}=0} {\rm Im}[\kappa({\bf k},{\bf p},{\bf q}) A({\bf p},t) A({\bf q},t ) A({\bf k},t)]\,.
\end{align}

Energy conservation requires that 
\begin{equation}
\frac{\partial [u_j({\bf k}, t) u^*_j({\bf k}, t)]}{\partial t} +\frac{\partial [u_j({\bf p}, t) u^*_j({\bf p}, t)]}{\partial t} +\frac{\partial [u_j({\bf q}, t) u^*_j({\bf q}, t)]}{\partial t} =0\,,
\end{equation}
which is equivalent to demanding
\begin{equation}
 \kappa({\bf k},{\bf p},{\bf q})+\kappa({\bf q},{\bf k},{\bf p})+ \kappa({\bf p},{\bf q},{\bf k})=0
\end{equation}
for any vectors ${\bf k}$, ${\bf p}$, and ${\bf q}$ satisfying ${\bf p}+{\bf q}+{\bf k}=0$. It is straightforward to check that the above relation is automatically satisfied given the expression of $\mathcal T$. Moreover, for $2+1$ fluids, by using the fact that ${\bf k} \cdot {\bf u}({\bf k}, t)={\bf p} \cdot {\bf u}({\bf p}, t)={\bf q} \cdot {\bf u}({\bf q}, t)=0$ and the identity
\begin{equation}
\sin^3\theta_1\cos(\theta_2-\theta_3)+\sin^3\theta_2\cos(\theta_3-\theta_1)+\sin^3\theta_3\cos(\theta_1-\theta_2)=0\,,
\end{equation}
\end{widetext}
for $\forall \theta_1+\theta_2+\theta_3=\pi\,,$ we can show that an additional symmetry for the mode-mode coupling exists, which is
\begin{equation}
k^2 \kappa({\bf k},{\bf p},{\bf q})+q^2\kappa({\bf q},{\bf k},{\bf p})+p^2 \kappa({\bf p},{\bf q},{\bf k})=0\,.
\end{equation}

This additional symmetry is directly connected with the additional conserved quantity in $2+1$ fluids: enstrophy. With two conserved quantities in the inertial regime, Kraichnan \cite{Kraichnan1967} explained that a dual-cascading behavior should be expected in the turbulent regime. This example strongly
suggests that the symmetry of the mode-mode coupling coefficients in our coupled oscillator model could be crucial for classifying the nonlinear behavior of gravitational evolutions.

\section{Expansion in two different bases}\label{sec:TwoBasis}

Let us imagine a simple example of a scalar field whose perturbations
propagate on a 2-dimensional flat spacetime with time-like boundaries
at $x=0$ and $x=1$. For comparison purposes, we have assigned two
coordinate systems in this spacetime: standard Cartesian coordinates
$(t, x)$ and ``null'' $(v, x)$ coordinates, with $v \equiv t+x$. For
simplicity, we impose Dirichlet boundary conditions
$\Phi |_{x=0}=\Phi |_{x=1} =0$ for the wave. At linear order, the
scalar wave satisfies the following wave equation
\begin{equation}
(-\partial^2_t+\partial^2_x)\Phi=0\,, 
\end{equation}
in the $(t, x)$ coordinate system or
\begin{equation}
(\partial^2_x+2\partial_v\partial_x)\Phi=0\,,
\end{equation}
in the $(v, x)$ coordinate system. 

Based on the wave equation and the boundary conditions, we can see
that this is a standard Sturm-Liouville problem, where it is
straightforward to write down the solutions of the wave equation in a
mode expansion
\begin{equation}
\Phi(t,x) = \sum_j \left(A_j e^{- i \omega_j t} +B_j e^{i \omega_j t}\right) \sin (j\pi x)\,,
\end{equation}
and 
\begin{equation}
\Phi(v,x) = \sum_j \left(\tilde A_j e^{- i \omega_j v} e^{i \omega_j x} +\tilde B_j e^{i \omega_j v} e^{- i \omega_j x}\right) \sin (j\pi x) \,,
\end{equation}
with $\omega_j =j \pi$. It is obvious that we can match up the linear
modes from the two different expansions above, and in fact we can make
the identifications
\begin{equation}\label{eqma}
A_j = \tilde A_j,\quad B_j =\tilde B_j\,.
\end{equation}

\begin{figure}[t,b]
\includegraphics[width=0.75\columnwidth]{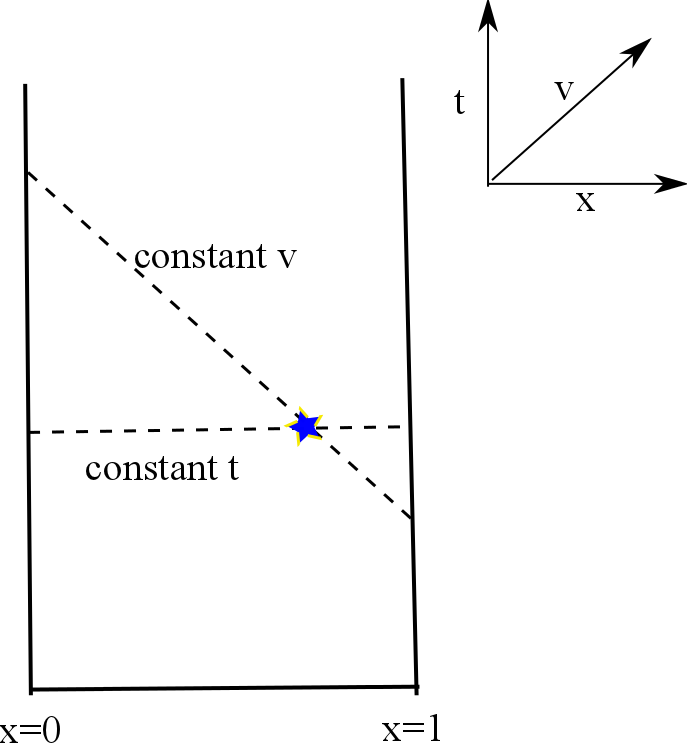}
\caption{An illustration for mode decompositions of a scalar field in
  a flat spacetime. At each point (such as the star in the diagram),
  we show two possible mode bases with respect to which to decompose
  the scalar wave.}
\label{fig:spacetime}
\end{figure} 

Now suppose nonlinear terms ($\Phi^2$, $\Phi^3$ or
even higher order) are present in the wave equations, resulting in a new solution
[$\Phi(t,x)$ or $\Phi(v,x)$]. For such a wave, we can still choose
constant-$t$ or constant-$v$ slices, and use the above spatial mode
basis to perform a decomposition
\begin{equation}
\Phi(t,x) = \sum_j \left[A_j(t) e^{- i \omega_j t} +B_j(t) e^{i \omega_j t}\right] \sin (j\pi x)\,,
\end{equation}
in the $(t, x)$ coordinates, and 
\begin{equation}
\Phi(v,x) = \sum_j \left[\tilde A_j(v) e^{- i \omega_j v} e^{i \omega_j x} +\tilde B_j(v) e^{i \omega_j v} e^{- i \omega_j x}\right] \sin (j\pi x)
\end{equation}
in the $(v, x)$ coordinates. We note that the mode amplitudes are
generically time-dependent now.

Pick an arbitrary point in the spacetime (for example, the one labeled
with a ``star'' in Fig.~\ref{fig:spacetime}). There we can ask whether
the matching described in Eq.~\eqref{eqma} still holds for the two
different mode expansions at that point. As we can see from
Fig~\ref{fig:spacetime}, these two mode expansions sample two
different slices of the spacetime: one at constant $t$ and the other
at constant $v$. Unlike the linear case, the scalar wave distributions
on these two slices can be made quite ``independent'' of each other by
freely detuning the nonlinear terms in the wave equations. In the end,
the largely independent data on these two slices imply that simple
mappings such as Eq.~\eqref{eqma} no longer exist for mode expansions
under different bases in the general nonlinear scenario. However, we
emphasize that despite the lack of a simple mapping between them, both
mode expansions are equally valid in describing the wave
evolution. Although our present analysis is performed using this
simple example where the mode expansion is complete, we see no reason
why a similar conclusion would not hold for quasinormal mode
expansions of generic spacetimes.

\section{Coupled oscillator model in Schwarzschild spacetime}\label{sec:schwarz}

As discussed in Sec.~\ref{sec2}, generic linear
metric perturbations can be decomposed into quasinormal modes plus a
residual part. Unless we are dealing with normal modes which form a
complete basis, or under certain physical conditions in which
quasinormal modes dominate (e.g., AdS perturbations in the
hydrodynamical limit), ignoring the contribution from the residual
part should always require justification. Here we offer an alternative
way of arriving at the coupled oscillator model, using the Green's
function approach (see also~\cite{Barranco:2013rua}). Using this
method, the quasinormal mode excitations can be unambiguously
determined given a driving source term. So far this approach can only
be demonstrated for perturbations with separable wave equations, such
as Schwarzschild and Kerr perturbations, and we shall leave extensions
to more general spacetimes to future studies.

To simplify the problem, we assume that the angular dependence has
been factored out, and we focus on the nonlinear evolution of modes
with spherical harmonic indices $(l, m)$, which satisfy the
Regge-Wheeler (odd partity) and Zerilli-Moncrief (even parity) wave
equations
\begin{equation}
\left [ -\frac{\partial^2}{ \partial t^2}+\frac{\partial^2}{\partial r^2_*}+V_{\rm e/o}(r )\right ] \Psi_{\rm e/o}=S_{\rm e/o}(r,t)\,.
\end{equation} 
Here $r_* \equiv r+ 2M \log[r/(2M)-1]$ and
$\Psi_{\rm e}, \Psi_{\rm o}$ are the Zerelli-Moncrief and Regge-Wheeler
gauge invariant quantities, respectively. The expressions for the
potential $V_{\rm e/o}$ and angular-projected source $S_{\rm e/o}$ can
be found in \cite{Martel2005,Yang:2014ae}. In our present
study, $S_{\rm e/o}$ is defined by the second order Ricci tensor,
which is bilinear in the metric perturbations.

Without the source term, for fixed time dependence $e^{-i \omega t}$
there are two independent solutions to each wave equation. One
solution asymptotes to
\begin{equation}
u_{\rm in} \rightarrow e^{-i \omega (t+r_*)},\quad r_* \rightarrow -\infty
\end{equation}
near the event horizon, and
\begin{equation}
u_{\rm in} \rightarrow C_{\rm in}(\omega) e^{-i \omega (t+r_*)}+C_{\rm out}(\omega)e^{-i \omega(r-r_*)},\quad r_* \rightarrow \infty
\end{equation}
at spatial infinity. The other solution satisfies 
\begin{equation}
u_{\rm out} \rightarrow e^{-i \omega (t-r_*)},\quad r_* \rightarrow \infty
\end{equation}
at the spatial infinity, and
\begin{equation}
u_{\rm out} \rightarrow \tilde{C}_{\rm in}(\omega) e^{-i \omega (t+r_*)}+\tilde{C}_{\rm out}(\omega)e^{-i \omega(r-r_*)},\quad r_* \rightarrow -\infty
\end{equation}
near the horizon. At the quasinormal mode frequencies $\omega_n$,
these two solutions become degenerate, and
$C_{\rm in}(\omega_n)=\tilde{C}_{\rm out}(\omega_n)=0$.

Using the Green's function technique, Leaver \cite{Leaver1986} showed
that $\Psi$ can be decomposed as
\begin{equation}
\Psi = \Psi_{\rm QNM} + \Psi_{\rm F}+\Psi_{\rm BC}\,,
\end{equation}
where $\Psi_{\rm F}$ is the contribution from high-frequency
propagator, $\Psi_{\rm BC}$ is the branch-cut contribution in the
Green function calculation, and $\Psi_{\rm QNM}$ is the quasinormal
mode contribution that we seek. In addition, he showed that
\begin{eqnarray}\label{eqgreen}
\Psi_{\rm QNM}(r,t) &= &2{\rm Re}\left [ \sum_n \frac{u_{\rm in}(r )e^{-i \omega_n t}}{D_n} \int^t_{-\infty}d t' \int^\infty_{-\infty}d r_*' \right . \nonumber \\
&&\quad\left . e^{i \omega_n t'} u_{\rm in}(r' ) S(r', t')\right ]\,,
\end{eqnarray}
with
\begin{equation}
D_n \equiv 2 \omega_n \left .\frac{d C_{\rm in}}{d \omega} \right |_{\omega_n}C^{-1}_{\rm out} (\omega_n)\,.
\end{equation}
Notice that we are taking the real part because this QNM contribution
is supposed to sum over both positive and negative frequencies. Also
note that in order to maintain causality, we have introduced an upper
bound $t$ into the time integral of Eq.~\eqref{eqgreen}, while in the
original paper \cite{Leaver1986} this bound was set to $\infty$ (see
also~\cite{Andersson:1996cm}). From Eq.~\eqref{eqgreen}, it is then
straightforward to derive the equations of motion for the amplitude
of mode $n$
\begin{equation}
\dot A_n(r, t) = \frac{e^{i \omega_n t}}{D_n} \int dr'_* u_{\rm in}(r') S(r', t) \equiv \frac{e^{i \omega_n t}}{D_n} \langle u_{\rm in} | S \rangle_{\text{CI}}\,, 
\end{equation}
where the integration should be performed as a contour integral in the
complex $r'$ plane to ensure convergence
\cite{Yang:2014zva}. Interestingly, when we apply this Green's
function technique to analyze generation of the shear quasinormal
modes in Sec.~\ref{sec4} (as the wave equation is separable), we find
that the generalized inner product $\langle\cdot |\cdot \rangle_{\rm CI}$ coincides
with $\langle\cdot |\cdot \rangle_\omega$ defined in
Eq.~\eqref{eq:InnerProd2}.

\bibliography{References}

\end{document}